\documentclass[usenatbib]{mn2e}
\usepackage{amsmath,amssymb}
\usepackage{graphicx}
\title[Introducing ambipolar diffusion into an AMR code]{Simulating hydromagnetic processes in star formation: introducing ambipolar diffusion into an adaptive mesh refinement code}
\author[D. F. Duffin and R. E. Pudritz]{D. F. Duffin$^{1}$\thanks{E-mail:
duffindf@mcmaster.ca} and R. E.
Pudritz$^{1,2}$\thanks{E-mail:
pudritz@physics.mcmaster.ca}\\
$^{1}$Department of Physics and Astronomy, McMaster University, Hamilton, Ontario L8S 4M1, Canada\\
$^{2}$Origins Institute, McMaster University, Arthur Bourns Bldg 241, Hamilton, Ontario L8S 4M1, Canada}
\begin{document}

\defcitealias{FM1993b}{FM93}

\date{Current Draft September 24 2007}

\pagerange{\pageref{firstpage}--\pageref{lastpage}} \pubyear{2008}

\maketitle

\label{firstpage}

\begin{abstract}

Given the importance of simulating hydromagnetic processes that impact star formation,  we have earlier developed a 3D adaptive mesh approach that allows us to include hydromagnetic processes during the formation and evolution of cores, discs, and stars in observed regions of star formation.  
In this paper, we take the next step in this program -- namely -- to develop a modified version of the 3D adaptive mesh refinement (AMR) code {\sc FLASH} \citep{2000ApJS..131..273F} in which the ambipolar diffusion of the magnetic field in poorly ionized molecular gas is implemented. 
We approach the problem using a single--fluid approximation to simplify numerical calculations.
In this paper, we present a series of test cases including oblique isothermal and non--isothermal C--shocks. We also present a study of the quasi--static collapse of an initial uniform, self--gravitating, magnetized sphere that is initially supported by its magnetic field against collapse (i.e. magnetically subcritical).  Applications to the collapse of a pre--stellar Bonnor--Ebert sphere are presented in a companion paper.  

\end{abstract}

\begin{keywords}
MHD -- shockwaves -- methods: numerical -- stars: formation -- ISM: clouds -- ISM: magnetic fields.
\end{keywords}

%%%%%%%%%%%%%%%%%%%%%%%%%%%%%%%%%%%%%%%%%%%%%%%%%%%%%%%%%%%%%%%%%%%%%%%%%%%%%%%%
%%%%%%%%%%%%%%%%%%%%%%%%%%%%%%%%%%%%%%%%%%%%%%%%%%%%%%%%%%%%%%%%%%%%%%%%%%%%%%%%
%%%%%%%%%%%%%%%%%%%%%    Introduction    %%%%%%%%%%%%%%%%%%%%%%%%%%%%%%%%%%%%%%%
%%%%%%%%%%%%%%%%%%%%%%%%%%%%%%%%%%%%%%%%%%%%%%%%%%%%%%%%%%%%%%%%%%%%%%%%%%%%%%%%
%%%%%%%%%%%%%%%%%%%%%%%%%%%%%%%%%%%%%%%%%%%%%%%%%%%%%%%%%%%%%%%%%%%%%%%%%%%%%%%%
\section{INTRODUCTION}
A powerful new theoretical picture of star formation, known as turbulent fragmentation, has emerged over the last decade. It emphasizes the role of supersonic turbulence in producing the filamentary structure and mass spectrum of overdense regions in Giant Molecular Clouds (GMCs) that ultimately collapse to form clusters of stars. A large number of simulations from a number of different groups have played a decisive role in the evolution of these ideas \citep[e.g. reviews by ][\citealt*{2007prpl.conf..149B}]{2004ARA&A..42..211E,2004RvMP...76..125M,2007prpl.conf..361A,2007prpl.conf...99K}. It has been demonstrated that 3D turbulence provides a reasonably good explanation of the origin of the core mass function (CMF). Cores are the overdense regions in which the `microphysics' of star formation \citep{2007arXiv0707.3514M} -- the formation of single or binary stars by gravitational collapse -- plays itself out. Fits of numerical models to observed CMFs are encouraging \citep{1999ApJ...525..318P}. Another important aspect of this picture is that the vorticity that is generated in such oblique shocks naturally explains the origin of angular momentum and formation of accretion discs observed around all young stars \citep[e.g. ][]{2004A&A...423....1J, 2004MNRAS.353..769T}. The subsequent fragmentation of discs depends both on their mass and angular momentum and is important for the genesis of binary stellar systems \citep{1964ApJ...139.1217T}. Only with the advent of fully 3D simulations has it been possible to move beyond highly idealized spherical or axisymmetric modeling to investigate non--equilibrium structures that bear close resemblance to those that are observed.

Magnetic fields can significantly affect these processes on both the macroscopic and microscopic scales of star formation. Although only a few 3D numerical hydromagnetic simulations are available, the results show that if molecular clouds have magnetic energy densities close to gravitational self--energy, fragmentation into stars can be strongly affected.  The so--called mass--to--flux ratio, $\Gamma \equiv M/\Phi = 2 \pi G^{1/2} \Sigma/B $, controls the evolution of clouds in in the limit that their magnetic fields are strongly coupled to the gas (`ideal magnetohydrodynamics (MHD)'). For values of $\Gamma \simeq 1-3$, magnetic fields force the collapse of structure into pancakes and can strongly limit the fragmentation into cores \citep[e.g. ][]{2007MNRAS.382...73T}. Robust fragmentation of overdense regions within GMCs into CMFs requires that they are sufficiently magnetically supercritical ($\Gamma > 1$) \citep*{2001ApJ...547..280H,2001ApJ...546..980O}. It is important to realize, however, that the turbulent fragmentation of  supercritical clouds still produces cores that have a distribution of magnetizations, most of which are more flux loaded (lower mass to flux ratio) than their large scale surroundings. In fact, even a strongly supercritical cloud can produce some cores that are critically magnetized \citep{1999ApJ...526..279P,2007MNRAS.382...73T}. This agrees with the observed presence of dynamically significant magnetic fields in a significant number of cores \citep[but not all;][]{1999ApJ...520..706C}.  

The strength of these magnetic effects can be much reduced, in principle, if fields are initially very weak or they are poorly coupled to the denser gas. The high column density of molecular gas implies that the gas--field coupling is far from perfect \citep{1981PASJ...33..617U}. In the bulk of the gas in molecular clouds, we observe ionization abundances on the order of $(10^{-7}) n_\mathrm{H}$. At the average densities in cluster forming gas within molecular clouds (e.g. densities of $\simeq 10^4$ cm$^{-3}$), the ionization fraction of molecular gas is lower but still leaves the field somewhat coupled and active in the dynamics. 
Only the charged species in a gas are directly coupled to the field, and it is through elastic collisions with neutral particles that the gas as a whole responds to magnetic forces.
Provided ion--neutral collision rates are large enough \citep*{2006ApJ...653.1280L}, charged species can attain significant velocities differing from the neutral gas velocity as they are dragged by the magnetic field lines in a process called ambipolar diffusion (AD).  
This process is also known as ambipolar drift, or as ion--neutral drift by plasma physicists.
With increasing slippage, this process can generate significant amounts of thermal energy -- more so than cosmic ray heating \citep*{2000ApJ...540..332P} -- and can act as a significant mechanism for the damping of turbulence in molecular clouds. 
In order to understand some of the effects of this complicated process, it is helpful to think of AD as lying somewhere in between the purely hydrodynamic (HD) behaviour of gas, and the perfectly coupled limit in which magnetic forces can strongly reshape purely HD behaviour.

The modeling of AD has been an important ingredient in star formation theory for decades \citep*{MS1956,1987ARA&A..25...23S}. This process was first used in an astrophysical context by \citet{MS1956} to describe the effective loss of magnetic support in a magnetically subcritical ($\Gamma < 1$) cloud of gas that is otherwise thermally supercritical, leading to its eventual collapse. Many early theoretical treatments of AD in the literature involve idealized one or two dimensional treatments in which only the evolution of non--rotating spheres or cylinders are treated. The role of ambipolar diffusion has been studied in the context of quasi--static collapse of magnetically supported \emph{thin discs} \citep{FM1993b,1994ApJ...432..720B,CB2006}, allowing for simplified geometries. This has recently been advanced by the correlated evolution of chemical species in the collapse \citep*{DM2001,2002ApJ...573..199N,TM2007b}.  

Simulations using the 3D Zeus--MP static grid code have studied the structure and stability of magnetic shocks in the presence of strong ambipolar diffusion \citep[C--shocks;][]{1994ApJ...425..171T,1995ApJ...442..726M,1997A&A...326..801S,1997ApJ...491..596M,2006ApJ...653.1280L}, including the first 3D simulations with ambipolar diffusion \citep{1997ApJ...491..596M}. 
\citet{1995ApJ...442..726M} started by implementing a single--fluid approximation, later developing a two--fluid model in order to capture shock instabilities \citep{1997ApJ...491..596M}.
Single--fluid approaches have been used to study star--formation and astrophysical turbulence \citep{HW2004a,2000ApJ...540..332P,1997ApJ...478..563Z,FM1993b}.
Multi--fluid codes have recently been developed to study the effects of ambipolar diffusion on turbulence and the formation of star forming clumps and shocks \citep{2006ApJ...638..281O,2006ApJ...653.1280L}, including more generalized physics \citep{2003MNRAS.344.1210F,2007MNRAS.376.1648O,2006MNRAS.366.1329O} all of which implement methods that avoid small timesteps.

In this and a series of upcoming papers, we investigate critical processes in star formation taking the effects of ambipolar diffusion into account through the use of the single--fluid approximation (carried out in \S\ref{sec:approx1}). Given certain physical conditions present in star--forming regions, we can ignore the charged fluid's inertia and pressure in a gas. This allows one to then treat only the dynamics of a single neutral fluid, with additional effects of the magnetic field due to elastic collisions with the charged fluid.  This  assumption is typically  made for the simplicity in implementing the approach and in the numerical calculations. We emphasize that this approximation is physically valid in star forming regions where we have low ionization and strong coupling of ions with the field.  
However, it must be noted that in small regions in molecular clouds this method is expected to be limited.  As an example, the small central regions within collapsing cores have densities exceeding $n_n > 10^{10}~\mathrm{cm^{-3}}$ such that grains will play a dynamical role.  This is also true in C--shocks (shocks mediated by ambipolar diffusion), where the width of the shock will be affected by the presence of grains.  The single--fluid approximation and its limitations are further discussed in \S\ref{sec:ions}, where we show that it is still a useful approximation.

In previous work, \citep{BP2006,BP2007} have successfully implemented the {\sc FLASH} AMR code to study the effects of ideal MHD in the context of star formation. In this paper we take the first step and outline the construction, testing, and behaviour of a 3D MHD code with adaptive mesh refinement (AMR) that employs ambipolar diffusion. Our code is an extension of the {\sc FLASH} code \citep{2000ApJS..131..273F}, and is thus general enough to treat many if not all of the problems mentioned above. Our code also treats the full energy equation  and is thus able to model effects such as drift heating and re--sorting of energy due to the diffusion of the field (important as a source of energy dissipation). The added advantage of AMR is particularly essential in the study of star--forming clumps which we study in some detail in our second paper. 

The extension of AD studies to three dimensions together with careful treatment of gravitational forces is an important step, for several reasons. First, star forming cores do not form in isolation in uniform media -- they are associated with filaments presumably created by systems of supersonic shocks \citep{2007prpl.conf..361A}. Second, cores have associated velocity gradients that can be interpreted as rotation \citep{2002ApJ...572..238C,2002ApJ...565..331C} and often (but not always) are threaded by fields whose mass--to--flux ratio is near unity \citep{2005LNP...664..137H}. The collapse of rotating objects that are threaded by magnetic fields leads to the creation of toroidal magnetic fields that play a central role in launching outflows \citep{BP2006,2007arXiv0705.2073M,2008A&A...477....9H}. However, toroidal fields and rotation have been left out of most simplifying models of AD and core evolution, or not fully implemented \citep{FM1993b,1994ApJ...432..720B,DM2001,CB2006,TM2007b}. Third, on the level of disc evolution, bars and spiral arms in discs can be very important in transporting disc angular momentum -- which can be competitive with extraction of angular momentum by magnetic tower flows and disc winds \citep{BP2006}. The weakened coupling of magnetic fields in the dense interiors of discs -- where AD will be relatively strong -- implies that angular momentum extraction may only be significant in disc surface layers as compared with transport by bars which could occur in the bulk of the material. Fourth, fragmentation of discs into binaries may be controlled by the strength and orientation of disc magnetic fields \citep{BP2006,2007MNRAS.377...77P}.

In the subsequent sections we first describe our implementation of ambipolar diffusion in the {\sc FLASH} code 
under the single--fluid approximation, including a brief discussion of it's limitations (\S 2). 
Next, we outline code tests (\S 3) that we have performed on isothermal and non--isothermal C--shocks. Our analysis provides a new important test for ambipolar diffusion codes which accurately account for heating and cooling (e.g. using a cooling look--up table instead of approximating the equation of state as a step function of density), and analytical solutions are developed. We then explore the combined effect of gravity, pressure, and magnetic fields with AD through a simplified version of the quasi--static collapse of a uniform, initially magnetically subcritical, gravitating sphere (\S 4).   We compare the results for pure hydro, ideal MHD, and our AD MHD and find interesting differences in the radial density profiles, which agree well with other related published results. We show that our code performs very well in difficult test problems, and is therefore an excellent tool with which to explore the full range of hydromagnetic processes that are relevant to stellar formation, binary formation, and perhaps even the formation of massive planetary systems by such processes.

%%%%%%%%%%%%%%%%%%%%%%%%%%%%%%%%%%%%%%%%%%%%%%%%%%%%%%%%%%%%%%%%%%%%%%%%%%%%%%%%
%%%%%%%%%%%%%%%%%%%%%%%%%%%%%%%%%%%%%%%%%%%%%%%%%%%%%%%%%%%%%%%%%%%%%%%%%%%%%%%%
%%%%%%%%%%%%%%%%%%%%%    Methodology     %%%%%%%%%%%%%%%%%%%%%%%%%%%%%%%%%%%%%%%
%%%%%%%%%%%%%%%%%%%%%%%%%%%%%%%%%%%%%%%%%%%%%%%%%%%%%%%%%%%%%%%%%%%%%%%%%%%%%%%%
%%%%%%%%%%%%%%%%%%%%%%%%%%%%%%%%%%%%%%%%%%%%%%%%%%%%%%%%%%%%%%%%%%%%%%%%%%%%%%%%

\section{METHODOLOGY}\label{sec:methodology}
As a general comment, our approach is to express the equations of partially ionized, magnetized, and self--gravitating gas is such a way so as to explicitly allow $\bmath{\nabla\cdot B} \ne 0$. We adopted this method because our numerical scheme ({\sc FLASH}) does not obey flux conservation, but instead, operates by keeping values below truncation--level error.  As will be discussed in a subsection below  (\S\ref{sec:divb}), we found after much experimentation that by allowing for non--zero values of the divergence of  the magnetic field in the equations, and then allowing the divergence cleaning procedure within the code to remove non--zero values (to below truncation error) after each time step, that greater accuracy and numerical stability was ensured for our code.  

The potential generality of our code was subjected to scrupulous testing. Analytical solutions to problems involving ambipolar diffusion are not particularly easy  given the fact that terms effectively contain second order derivatives in the single--fluid approximation (or involve two or more interacting fluids in the multiple fluid picture).  Common tests include the formation of stable isothermal C--shocks \citep{1995ApJ...442..726M,2003MNRAS.344.1210F,2006MNRAS.366.1329O,2006ApJ...653.1280L}, and performing a similar quasi--static collapse to that of \citet{FM1993b} \citep*{SMS1997,HW2004a}.  

The former is an excellent test with analytic solutions, although isothermal.  We present herein an adaptation of the C--shock test which allows for testing of a code with ambipolar diffusion energy terms.  This differs only slightly, though critically, from the non--isothermal analytical solution of \citep{1991MNRAS.251..119W}.  The collapse test is a more qualitative test and involves measuring radial profiles and the time--scale of collapse when AD is present. We note that we have already rigorously tested our MHD module in several other papers in the context of magnetized collapse problems in both HD and ideal MHD.    

%%%%%%%%%%%%%%%%%%%%%%%%%%%%%%%%%%%%%%%%%%%%%%%%%%%%%%%%%%%%%%%%%%%%%%%%%%%%%%%%
%%%%%%%%%%%%%%%%%%%%%%%%%%%%%%%%%%%%%%%%%%%%%%%%%%%%%%%%%%%%%%%%%%%%%%%%%%%%%%%%
%%%%%%%%%%%%%%%%%%%%%    Theory of AD    %%%%%%%%%%%%%%%%%%%%%%%%%%%%%%%%%%%%%%%
%%%%%%%%%%%%%%%%%%%%%%%%%%%%%%%%%%%%%%%%%%%%%%%%%%%%%%%%%%%%%%%%%%%%%%%%%%%%%%%%
%%%%%%%%%%%%%%%%%%%%%%%%%%%%%%%%%%%%%%%%%%%%%%%%%%%%%%%%%%%%%%%%%%%%%%%%%%%%%%%%
\subsection{Theory of ambipolar diffusion}\label{sec:adtheory}

Consider a gas composed of both ions (subscript i) and neutrals (subscript n).  Electrons present in the fluid are considered well coupled to the ions in the regime we wish to study.  The MHD  equations for this two--component system, in which only the ions couple to the field, are,
\begin{equation}\label{contn}
\frac{\partial\rho_{n}}{\partial t} + \bmath{\nabla} \bmath\cdot \left(\rho_{n} \bmath{u}_{n}\right) = 0 
\end{equation}
\begin{equation}\label{conti}
\frac{\partial\rho_{i}}{\partial t} + \bmath{\nabla} \bmath\cdot \left(\rho_{i} \bmath{u}_{i}\right) = 0
\end{equation}
\begin{equation}\label{consmomn}
\frac{\partial\left(\rho_{n}\bmath{u}_{n}\right)}{\partial t} +
\bmath{\nabla} \bmath\cdot
\left(\rho_{n}
\bmath{u}_{n}
\bmath{u}_{n} + P_{n}\right) =
- \rho_{n} \bmath{g} -
\bmath{f}_{f}
\end{equation}
\begin{equation}\label{consmomi}
\begin{split}
\frac{\partial\left(\rho_{i}\bmath{u}_{i}\right)}{\partial t} + &
\bmath{\nabla} \bmath\cdot
\left(\rho_{i}
\bmath{u}_{i}
\bmath{u}_{i} + P_{i} +
\frac{B^{2}}{2 {\mu}_{0}} -
\frac{1}{{\mu}_{0}}\bmath{B B}\right) =
- \rho_{i} \bmath{g} \\ & +
\bmath{f}_{f}-\frac{1}{{\mu}_{0}}\bmath{B}
\left(\bmath{\nabla} \bmath\cdot
\bmath{B}\right),
\end{split}
\end{equation}
where $\rho$ represents density, $\bmath{u}$ is a velocity, $P$ is a pressure, $\bmath{B}$ is the magnetic field, $\bmath{g}$ is the gravitational acceleration and $\bmath{f}_{f}$ is the frictional force density from ion--neutral collisions \citep{1978ppim.book.....S},
\begin{equation}\label{friction}
\bmath{f}_{f} \equiv \gamma_\mathrm{AD} \rho_{i} \rho_{n} \left(\bmath{u}_{n}-\bmath{u}_{i}\right)
= - \frac{1}{{\mu}_{0} \beta_\mathrm{AD}} \bmath{u}_{d}, 
\end{equation}
where $\beta_\mathrm{AD} = \frac{1.4}{{\mu}_{0} \gamma_\mathrm{AD} \rho_{i} \rho_{n}}$ and the drift velocity $\bmath{u}_{d}$ is defined as,
\begin{equation}\label{drift1}
\bmath{u}_{d} \equiv \bmath{u}_{i}-\bmath{u}_{n}.
\end{equation}

The constant $\gamma_\mathrm{AD} = \frac{< \sigma \omega>_{ni}}{m_{i} + m_{n}} = 3.28 \times 10^{13}\ \mathrm{g}^{-1}\ \mathrm{cm}^{3}\ \mathrm{s}^{-1}$ represents the coupling of the neutrals and ions (the drag coefficient).  Ions are considered to be typically HCO$^{+}$ or Na$^{+}$ which have similar masses (about 29.0 a.m.u.) and collision rates with H$_{2}$ $\nolinebreak{(<\sigma \nu>_\mathrm{ni}}= 1.7 \times 10^{-9}\ \mathrm{cm}^{-3}\ \mathrm{s}^{-1}$ \citep{1973ppim.book.....M}).  The value of 1.4 in $\beta_\mathrm{AD}$ arises from the fact that we have about 10\% {\sc He} per {\sc H} atom in our gas \citep{FM1993b,HW2004a} and is  1.0 in the absence of {\sc He} (such as in our C--shock test runs).  Helium is heavier than H$_2$, and thus changes the collisional dynamics of the neutral gas.

Equations (\ref{contn}) and (\ref{conti}) are the continuity equations of the neutrals and ions respectively, expressing conservation of mass.  Equations (\ref{consmomn}) and (\ref{consmomi}) are expressions of conservation of momentum.  Note that while the ions undergo direct MHD forces, the neutrals do so only by collisions with the ions through the friction term in the HD momentum equation (\ref{consmomn}).  
 
The induction equation -- coupled to the ions -- is,
\begin{equation}\label{induct1}
\frac{\partial\bmath{B}}{\partial t}=- \bmath{\nabla} \bmath\cdot \left(\bmath{u}_{i} \bmath{B - B u}_{i}\right) 
- \left(\bmath{\nabla \cdot B}\right) \bmath{u}_{i}.
\end{equation}
This defines the evolution of the magnetic field.

Expressing conservation of energy we find the rest of our initial equations,
\begin{equation}\label{enern}
\frac{\partial {E}_{n}}{\partial t} + \bmath{\nabla} \bmath\cdot \left[\bmath{u}_{n}\left(E_{n} + P_{n}\right)\right] =
-\bmath{u}_{n} \bmath\cdot \bmath{f}_{f} + {\rho}_{n} \bmath{g \cdot u}_{n}
\end{equation}
\begin{equation}\label{eneri}
\begin{split}
\frac{\partial {E}_{i}}{\partial t} + & \bmath{\nabla} \bmath\cdot \left[\bmath{u}_{i}\left(E_{i} + P_{i}
 + \frac{B^{2}}{2 \mu_{0}}\right)+ \frac{1}{{\mu}_{0}} \left(\bmath{u}_{i}  \bmath\cdot \bmath{B}\right) \bmath{B}  \right] = 
\bmath{u}_{i} \bmath\cdot \bmath{f}_{f}\\ & + {\rho}_{i} \bmath{g \cdot u}_{i} 
  -\frac{1}{{\mu}_{0}}\left(\bmath{u}_{i} \bmath\cdot \bmath{B}\right)\left(\bmath{\nabla \cdot B} \right), 
\end{split}
\end{equation}
where we see again that the neutrals undergo purely HD forces while the forces on the ions are MHD.  The energy densities are respectively defined as:
\begin{equation}
E_n = \frac{1}{2}\rho_n u_n^2 + \frac{1}{\gamma-1}P_n
\end{equation}
\begin{equation}
E_i = \frac{1}{2}\rho_i u_i^2 + \frac{1}{\gamma-1}P_i + \frac{1}{2\mu_0}B^2,
\end{equation}
where $\gamma$ is the adiabatic index of the gas (we consider equal indexes for both gases).  
%%%%%%%%%%%%%%%%%%%%%%%%%%%%%%%%%%%%%%%%%%%%%%%%%%%%%%%%%%%%%%%%%%%%%%%%%%%%%%%%
%%%%%%%%%%%%%%%%%%%%%%%%%%%%%%%%%%%%%%%%%%%%%%%%%%%%%%%%%%%%%%%%%%%%%%%%%%%%%%%%
%%%%%%%%%%%%%%%%%%%%%    Ionization Considerations    %%%%%%%%%%%%%%%%%%%%%%%%%%
%%%%%%%%%%%%%%%%%%%%%%%%%%%%%%%%%%%%%%%%%%%%%%%%%%%%%%%%%%%%%%%%%%%%%%%%%%%%%%%%
%%%%%%%%%%%%%%%%%%%%%%%%%%%%%%%%%%%%%%%%%%%%%%%%%%%%%%%%%%%%%%%%%%%%%%%%%%%%%%%%

\subsection{Ionization considerations and approximations}\label{sec:ions}
One can simplify these equations by noting that the ionization fraction in molecular clouds and protoplanetary discs is very low (${\rho}_{i} \ll {\rho}_{n}$).  
Subsequent conditions are that the ions are perfectly coupled to the field and that the ion--neutral collisional timescale (the timescale for an ion to collide with any neutral particle) time is much less than other physical timescales in our system  \citep[e.g.][]{2006ApJ...653.1280L} so that the fluids are sufficiently coupled.  This can help effectively eliminate many of the ion equations (see \S\ref{sec:approx1}). Although we will always be left with $\rho_{i}$ in the $\beta_\mathrm{AD}$ term which determines the strength of the ambipolar diffusion (lower ion density means stronger ambipolar diffusion).

This approach gives us the advantage of a scheme that decreases the number of numerical calculations compared to many-fluid schemes, in addition to a relatively simple implementation.  However, there are limitations to this approach.  Primarily, there are situations in star forming regions in which the assumptions break down.  These include small, dense regions in collapsing cores with densities that exceed $n_n>10^{10}~\mathrm{cm^{-3}}$, and in certain shocks.  When collisions heat up the flow in a shock so that the neutral flow becomes subsonic, the shock may undergo a short, sharp transition \citep[sub--shock;][]{1993ARA&A..31..373D}.  When this happens, the shock is called a J--shock.  The ion inertial equation is important in the formation of this structure, though schemes which neglect ion--inertia have been successful in modelling it \citep{2003MNRAS.344.1210F}.  This sub--shock structure will play an important role in the chemistry and heating of the shock.  Strong J--shocks are thought to be common in star--forming regions \citep{1987ApJ...312..143C}.

In C--shocks, where the transition of the shock is supersonic and everywhere continuous, the flow is susceptible to the Wardle instability \citep{1990MNRAS.246...98W}. This instability is dependent on the continuity equation of the ions, which is neglected under the single fluid approximation.  The instability -- caused by the buckling of field lines along the shock front -- leads to the formation of thin sheets (in 3D) with density enhancements of about $10^2$ times the ambient density. These thin sheets -- with thickness on the order of $l_{WI} = 10^{-2}~L_\mathrm{shock} \approx 2\times10^{13}~\mathrm{cm}$ in astrophysical simulations \citep{1997ApJ...491..596M} -- will most likely play a role physically in heating the shock and in creating energetic particles.  The sheets are too thin, however, to play a role in creating dense cores which could contribute to the CMF, as has been suggested by \citet{1994ApJ...425..171T}.  For example, by looking at the simulations of \citet{1997ApJ...491..596M} it can be estimated that the Jeans length of the thin sheets will be (using their parameters)
\begin{equation}
\lambda_J \approx 1.7\times10^{15} \left(\frac{c_s}{0.01~\mathrm{km~s^{-1}}}\right) \left(\frac{n_n}{10^7~\mathrm{cm^{-3}}}\right)^{-1/2}~\mathrm{cm}\gg l_{WI}.
\end{equation}
Thus, the Wardle instability will play an important role in mediating the radiation profiles of shocks, but not play as important a role for the larger scale dynamics that we wish to study.  

The limitation of the single fluid approximation can be expressed more explicitly for regions where the ambipolar Reynold's number \citep{2002ApJ...567..962Z}
\begin{equation}
R_\mathrm{AD} = \frac{u L_B}{\beta_\mathrm{AD} B^2} = \frac{u}{u_d} < 1,
\end{equation}
where $u$ is a typical velocity and $L_B$ is a typical lengthscale on which the field varies.  Under these circumstances the ions and neutrals will not be sufficiently coupled to allow the approximation \citep{2000ApJ...540..332P}, an important regime for the dissipation of astrophysical turbulence \citep{2006ApJ...638..281O}.  In astrophysical situations, it can be shown that $u > u_d$ is generally valid.  Following \citet{1997ApJ...478..563Z}, we assume $L_B\approx (10^{20}~\mathrm{cm^{-2}})/ n_n$ based on observed column densities of diffuse clouds.  This allows us to derive a value for  $u_d/u$ using equation \eqref{ionization} for the ionization, a typical flow velocity of $u\approx 1.5~\mathrm{km~s^{-1}}$ ($\mathcal{M}=5$) and a magnetic field strength of $10~\mathrm{\mu G}$. The relation satisfies $u_d/u<1$ for all $n_n$, and for the most part satisfies $u_d/u\ll 1$.  For example, $u_d/u = [6\times10^{-3}, 0.5, 3\times 10^{-2}, 3\times 10^{-4}]$ for neutral densities of $n_n = [10^2, 2.1\times 10^{3}, 10^{6}, 10^{10}]~\mathrm{cm^{-3}}$, where $n_n=2.1\times 10^{3}~\mathrm{cm^{-3}}$ gives the maximum value of $(u_d/u)_\mathrm{max}=0.5$.  Thus the approximation is a valid one on the scales we are interested in.

In this paper, we shall use a simplified approach in treating the ionization of molecular clouds which is pertinent to lower density regimes, and which has been used widely in the literature of the subject. The ion density can be expressed as a simple function of neutral density in this regime. We follow the lead of other authors \citep{FM1993b, HW2004b} and use the very simple expression (see the quasi--static collapse problem):  
\begin{equation}\label{ionization}
n_{i} = K\left(\frac{n_{n}}{10^{5}\ \mathrm{cm}^{-3}}\right)^{k} + K' \left(
\frac{n_{n}}{10^{3}\ \mathrm{cm}^{-3}}\right)^{-2}, 
\end{equation}
where $n$ is a number density, $K=3 \times 10^{-3}\ \mathrm{cm}^{-3}$, $k=\frac{1}{2}$ and $K'=4.64\times10^{-4}\ \mathrm{cm}^{-3}$.  This approximation is meant for density situations in which dust grains do not play a substantial role in the ionization balance.  It allows one to eliminate entirely the need to track the ion density in the single fluid approximation.  The second term dies off quickly in the higher density regime, at which point we're left with the common $n_i \propto n_n^{1/2}$ relation. One can, by these means, reduce the 2--fluid (ions plus neutrals) to that of an effectively single fluid with low conductivity. We note that this approach is still very important and useful since it allows one to accurately track the physics of core formation and collapse up to densities that far exceed the density of cores that are nearly in equilibrium (at $10^4 - 10^6$ cm$^{-3}$), which is our intention.

This well--known expression arises by approximating the results of ionization equilibrium calculations \citep{1979ApJ...232..729E, 1979PASJ...31..697N} where the sole form of ionization is through cosmic rays (where the cited authors assume a cosmic ray ionization rate of $\zeta_0 = 6.9 \times 10^{-17}\ \mathrm{s}^{-1}$).  It is approximately correct for densities $n< 10^{10}\ \mathrm{cm}^{-3}$, after which charged grains become the principal charge carriers \citep{TM2007b}.  Also, more active and complicated chemistry will be required to model C-shocks \citep{1986MNRAS.220..801P}.  However, for our purposes, it is sufficient that it matches the equation originally used by \citet{FM1993b}. 
 Furthermore, in order to study the collapse and disc formation to densities greater than $10^{10}$  cm$^{-3}$ a multi--fluid approach including grains will be required.

For the C--shock test we use a constant ion density.  This leads to slightly different analytic solutions than used in the past.  We demonstrate how to adapt given a certain ion density prescription, or non at all (ion mass conservation).    

%%%%%%%%%%%%%%%%%%%%%%%%%%%%%%%%%%%%%%%%%%%%%%%%%%%%%%%%%%%%%%%%%%%%%%%%%%%%%%%%
%%%%%%%%%%%%%%%%%%%%%%%%%%%%%%%%%%%%%%%%%%%%%%%%%%%%%%%%%%%%%%%%%%%%%%%%%%%%%%%%
%%%%%%%%%%%%%%%%%%%%%%%%%%%%%   One Fluid    %%%%%%%%%%%%%%%%%%%%%%%%%%%%%%%%%%%
%%%%%%%%%%%%%%%%%%%%%%%%%%%%%%%%%%%%%%%%%%%%%%%%%%%%%%%%%%%%%%%%%%%%%%%%%%%%%%%%
%%%%%%%%%%%%%%%%%%%%%%%%%%%%%%%%%%%%%%%%%%%%%%%%%%%%%%%%%%%%%%%%%%%%%%%%%%%%%%%%
\subsection{Carrying out the single fluid approximation}\label{sec:approx1}
Our goal in this Section is to make equations (\ref{contn}), (\ref{consmomn}), (\ref{induct1}) and (\ref{enern}) look like MHD equations for the neutrals plus some other terms. This allows us to simplify the computations significantly by reducing the number of equations that must be solved at each time step\footnote{More general and robust multi--fluid techniques have been developed \citep{2007MNRAS.376.1648O,2006MNRAS.366.1329O,2006ApJ...653.1280L,2006ApJ...638..281O,2003MNRAS.344.1210F}, although isothermal and on a static grid.  These techniques may be more important in turbulent conditions where pertinent timescales are highly variable.}  .  We have already noted that molecular clouds are  dense and poorly ionized which implies that ${\rho}_{i} \ll {\rho}_{n}$.  Furthermore, we require that the region we study have an ion--neutral collisional time ($t_\mathrm{in} = 1/(\gamma_\mathrm{AD}\rho_\mathrm{n})$) that is much shorter than other physical timescales in the problem  \citep{2006ApJ...653.1280L}.  This is also typically true in molecular clouds, which allows us to ignore the ion inertial forces with respect to their magnetic and frictional forces.  As is well known, one can use this to reduce the ion equations of motion into one equation for the drift velocity, saving half the calculations.  

More explicitly, we find that the gravitational and pressure forces for the ions are negligible with respect to the magnetic force $\bmath{f}_m = \bmath{J \times B}$ (where $ \bmath{J} = \frac{1}{{\mu}_{0}} \bmath{\nabla \times B}$).  We can then ignore the ion inertia through our secondary approximation.  Equation (\ref{consmomi}) leaves us the equation of force balance -- between the full Lorentz force exerted directly upon the ions, and the friction force that arises from ion--neutral collisions:
\begin{equation}\label{fffm}
\bmath{f}_f = -\bmath{f}_m=-\bmath{J \times B}, 
\end{equation}
from which we can solve for the velocity of the ions:
\begin{equation}\label{drift2}
\bmath{u}_{d} = \beta_\mathrm{AD} (\bmath{\nabla \times B}) \times \bmath{B}
\end{equation}
\begin{equation}\label{drift3}
\bmath{u}_{i} = \bmath{u}_{n} + \beta_\mathrm{AD} (\bmath{\nabla \times B}) 
\times \bmath{B}.
\end{equation}

Substituting (\ref{drift3}) into (\ref{induct1}) we can solve to get [simplifying $\bmath{\nabla \cdot (uB - Bu)} = -\bmath{\nabla \times (u \times B)}$],
\begin{equation}\label{induct2}
\begin{split}
\frac{\partial\bmath{B}}{\partial t} = & \bmath{\nabla} \times \left(\bmath{u}_n \times \bmath{B}\right) - \left(\bmath{\nabla \cdot B}\right)\bmath{u}_{n}\\ & + \bmath{\nabla} \times \left\{\beta_\mathrm{AD}\left[\left(\bmath{\nabla \times B}\right)\times\bmath{B}\right]\times\bmath{B}\right\}\\
& - \left(\bmath{\nabla \cdot B}\right)\left[\beta_\mathrm{AD}\left(\bmath{\nabla \times B}\right)\times\bmath{B}\right],
\end{split}
\end{equation}
where the neutrals look to be perfectly coupled to the magnetic field save for the two last terms which describe the ambipolar diffusion of the field.

Similarly, in the energy equation of the neutrals (\ref{enern}) we substitute into the frictional force (\ref{fffm}).   We recover,
\begin{equation}\label{energy1}
\begin{split}
\frac{\partial {E}_{n}}{\partial t} & + \bmath{\nabla} \bmath\cdot \left[\bmath{u}_{n}\left(E_{n} + P_{n}
 + \frac{B^{2}}{2 {\mu}_{0}}\right)+ \frac{1}{{\mu}_{0}} \left(\bmath{u}_{n}  \bmath\cdot \bmath{B}\right) \bmath{B}  \right]\\ &
+\bmath{\nabla}\bmath\cdot\left[\beta_\mathrm{AD} B^{2} \left( \bmath{J \times B}\right)\right] = {\rho}_{n} \bmath{g \cdot u_{n}}\\ &
-\frac{1}{{\mu}_{0}}\left(\bmath{u_{n} \cdot B}\right)\left(\bmath{\nabla \cdot B}\right)
+{\mu}_{0}\beta_\mathrm{AD}\parallel \bmath{J \times B} {\parallel}^{2}\\ & - \beta_\mathrm{AD}
\left[\bmath{B \cdot \left(J \times B\right)}\right]\left(\bmath{\nabla \cdot B}\right), 
\end{split}
\end{equation}
where the last term on the LHS and the last two terms of the RHS are ambipolar diffusion terms.  Heating of the fluid occurs through the dissipation of the field (${\mu}_{0}\beta_\mathrm{AD}\parallel \bmath{J \times B} {\parallel}^{2}$ term).  Diffusion of energy is similar to that of Ohmic diffusion ($\bmath{\nabla}\bmath\cdot\left[\beta_\mathrm{AD} B^{2} \left( \bmath{J \times B}\right)\right]$ term), given the ambipolar diffusivity below.  The rest (ignoring $\bmath{\nabla\cdot B}$ terms) looks like the MHD energy equation for the neutral species; the total energy of the fluid is changed by the magnetic field as it undergoes motion in the gravitational field.  

Now that the single fluid equations have been written out, we may dispense with the subscript n (i.e. representing the gas as a whole).  The following MHD equations are now in a form that is optimized for our computational 
approach (i.e. similar to \citet{1999JCP.154.284P}), wherein fluxes appear within a divergence on the LHS and sources are placed on the RHS:
\begin{equation}\label{continuity}
\frac{\partial \rho}{\partial t} + \bmath{\nabla} \bmath\cdot \left(\rho \bmath{u}\right) = 0
\end{equation}

\begin{equation}\label{momentum}
\begin{split}
\frac{\partial\left(\rho\bmath{u}\right)}{\partial t} + &
\bmath{\nabla} \bmath\cdot
\left(\rho \bmath{u} \bmath{u} + P +
\frac{B^{2}}{2 {\mu}_{0}} -
\frac{1}{{\mu}_{0}}\bmath{B B}\right) =
- \rho \bmath{g}\\ & - \frac{1}{{\mu}_{0}}\bmath{B}
\left(\bmath{\nabla} \bmath\cdot \bmath{B}\right)\\
\end{split}
\end{equation}

\begin{equation}\label{induction}
\begin{split}
\frac{\partial \bmath{B}}{\partial t} & + \bmath{\nabla} \bmath\cdot \left(\bmath{uB - Bu}\right)\\ & + \bmath{\nabla}
\bmath\cdot \left\{\mu_{0}\beta_\mathrm{AD}\left[ \bmath{\left(J \times B\right)B - B\left(J \times B\right)}\right]\right\}\\ & = -\left(\bmath{\nabla \cdot B}\right)\bmath{u} - \left(\bmath{\nabla 
\cdot B}\right)\left[\mu_{0}\beta_\mathrm{AD} \bmath{\left(J \times B\right)}\right]
\end{split}
\end{equation}

\begin{equation}\label{energy}
\begin{split}
\frac{\partial E}{\partial t} & + \bmath{\nabla} \bmath\cdot \left[\bmath{u}\left(E + P
 + \frac{B^{2}}{2 {\mu}_{0}}\right)+ \frac{1}{{\mu}_{0}} \bmath{\left(u  \cdot B\right) B }  \right]\\ & 
+\bmath{\nabla}\bmath\cdot\left[\beta_\mathrm{AD} B^{2} \left( \bmath{J \times B}\right)\right] = \rho \bmath{g \cdot u} -\frac{1}{{\mu}_{0}}\left(\bmath{u \cdot B}\right)\left(\bmath{\nabla \cdot B}\right)\\ &
+{\mu}_{0}\beta_\mathrm{AD}\parallel \bmath{J \times B} {\parallel}^{2}  - \beta_\mathrm{AD}\left[\bmath{B \cdot \left(J \times B\right)}\right]\left(\bmath{\nabla \cdot B}\right).
\end{split}
\end{equation}

%%%%%%%%%%%%%%%%%%%%%%%%%%%%%%%%%%%%%%%%%%%%%%%%%%%%%%%%%%%%%%%%%%%%%%%%%%%%%%%%
%%%%%%%%%%%%%%%%%%%%%%%%%%%%%%%%%%%%%%%%%%%%%%%%%%%%%%%%%%%%%%%%%%%%%%%%%%%%%%%%
%%%%%%%%%%%%%%%%%%%%%%%%%%%%%%%%   Timesteps    %%%%%%%%%%%%%%%%%%%%%%%%%%%%%%%%
%%%%%%%%%%%%%%%%%%%%%%%%%%%%%%%%%%%%%%%%%%%%%%%%%%%%%%%%%%%%%%%%%%%%%%%%%%%%%%%%
%%%%%%%%%%%%%%%%%%%%%%%%%%%%%%%%%%%%%%%%%%%%%%%%%%%%%%%%%%%%%%%%%%%%%%%%%%%%%%%%

\subsection{Diffusion and ambipolar drift:  typical timesteps}\label{sec:diffusion}
A glance at the form of the induction equation above \eqref{induction} clearly shows that  \emph{ambipolar diffusion (ion--neutral drift) is not just the simple diffusion of the field} \citep{2006ApJ...638..281O,1995ApJ...442..726M,1994ApJ...427L..91B}.  Ohmic diffusion, for instance, only adds a term of the form $\bmath{\nabla} \times (\eta \bmath{\nabla \times B})$ to the LHS of the induction equation (under a similar approximation), where $\eta$ is the Ohmic diffusivity.  The corresponding ambipolar diffusivity is given as \citep{2002ApJ...567..962Z}, 
\begin{equation}\label{addiffusivity}
\eta_{\mathrm{AD}} = \beta_\mathrm{AD} B^{2}, 
\end{equation}  
where $\eta_{\mathrm{AD}}$ has units of cm$^2\ \mathrm{s}^{-1}$.  This can be used to re--evaluate the ambipolar diffusion induction term as \citep{1994ApJ...427L..91B},
\begin{equation}\label{addiffusion}
\bmath{\nabla} \times \left[\eta_{AD}\bmath{\nabla \times B} - \mu_{0}\beta_\mathrm{AD}(\bmath{J \cdot B})\bmath{B}\right]
\end{equation}
on the LHS.  Clearly the second term demonstrates the deviation of ambipolar diffusion (ion--neutral drift) from a purely diffusive process.  

In fact, it is easy to imagine situations in which $\bmath{J \cdot B} \ne 0$, such as disc accretion in a collapsing core.  It would then not be surprising that the extra non--diffusive term becomes important as field lines are strongly oriented against the accretion flow, which caries a charged current.  Nonetheless, we expect the diffusive component to dominate.  We thus use a diffusive scaling to estimate the typical time--scale that is needed in order to satisfy the Courant condition \citep{1995ApJ...442..726M},
\begin{equation}\label{adtime--scale}
\tau_{\mathrm{AD}} = T_0 \frac{(\Delta x)^{2}}{\eta_{\mathrm{AD}}},  
\end{equation}
where the factor of $T_0$ is a fudge factor (we use $T_0 = 0.00833 \approx 1/120$ in our simulations as used by \citet{1995ApJ...442..726M}) and $\Delta x$ is a typical length scale (see \S\ref{sec:tad}).  Longer time--scales or slower speeds (if the timestep is not used, for instance) tend to allow sharp gradients in the field to form which lead to instabilities generated by the ambipolar diffusion terms.  This creates unphysical states such as negative densities.

%%%%%%%%%%%%%%%%%%%%%%%%%%%%%%%%%%%%%%%%%%%%%%%%%%%%%%%%%%%%%%%%%%%%%%%%%%%%%%%%
%%%%%%%%%%%%%%%%%%%%%%%%%%%%%%%%%%%%%%%%%%%%%%%%%%%%%%%%%%%%%%%%%%%%%%%%%%%%%%%%
%%%%%%%%%%%%%%%%%%%%%    The FLASH2.5 Code    %%%%%%%%%%%%%%%%%%%%%%%%%%%%%%%%%%
%%%%%%%%%%%%%%%%%%%%%%%%%%%%%%%%%%%%%%%%%%%%%%%%%%%%%%%%%%%%%%%%%%%%%%%%%%%%%%%%
%%%%%%%%%%%%%%%%%%%%%%%%%%%%%%%%%%%%%%%%%%%%%%%%%%%%%%%%%%%%%%%%%%%%%%%%%%%%%%%%
\subsection{The {\sc FLASH2.5} code}\label{sec:FLASH}
The {\sc FLASH} code \citep{2000ApJS..131..273F} employs a wide range of HD and MHD physics.  It has been well tested and has become quite popular in the astrophysical community.  Its principal advantage is that it can perform AMR, as this saves considerable computational time while maintaining a high level of refinement.  A secondary, though important advantage for astrophysicists, is that it does a very good job in capturing shocks.

The refinement criteria for {\sc FLASH} have been adjusted somewhat in previous work by \cite*{BPH2004}.  It was shown by \citet{1997ApJ...489L.179T} that one needs to refine the Jeans length, 
\begin{equation}\label{eqn:jeans}
\lambda_J = \left(\frac{\pi c_s^2}{G \rho}\right)^{1/2}, 
\end{equation}
by at least four cells in an AMR code, where $c_s$ is the isothermal sound speed, $G$ is Newton's constant and $\rho$ is the local density.  If this is not satisfied artificial fragmentation induced by the numerical grid will result.  \citet{BPH2004} setup the refinement criteria  such that a variable fraction of cells will resolve the Jeans length.  In the quasi--static collapse we use 8 cells per Jeans length, guaranteeing artificial fragmentation will be avoided. 

Our explicit coding of  the above single--fluid approximation in {\sc FLASH} is described in detail in Appendix (\ref{sec:mhdimp}).  This includes the splitting of the equations, the timestep condition and the dimensionality of the equations.  

%%%%%%%%%%%%%%%%%%%%%%%%%%%%%%%%%%%%%%%%%%%%%%%%%%%%%%%%%%%%%%%%%%%%%%%%%%%%%%%%
%%%%%%%%%%%%%%%%%%%%%%%%%%%%%%%%%%%%%%%%%%%%%%%%%%%%%%%%%%%%%%%%%%%%%%%%%%%%%%%%
%%%%%%%%%%%%%%%%%%%%%    DivB    %%%%%%%%%%%%%%%%%%%%%%%%%%%%%%%%%%%%%%%%%%%%%%%
%%%%%%%%%%%%%%%%%%%%%%%%%%%%%%%%%%%%%%%%%%%%%%%%%%%%%%%%%%%%%%%%%%%%%%%%%%%%%%%%
%%%%%%%%%%%%%%%%%%%%%%%%%%%%%%%%%%%%%%%%%%%%%%%%%%%%%%%%%%%%%%%%%%%%%%%%%%%%%%%%

\subsection{$\bmath{\nabla\cdot B}$ considerations}\label{sec:divb}
The MHD module in {\sc FLASH} employs a somewhat controversial scheme that does not explicitly constrain $\bmath{\nabla\cdot B}=0$, based on \citet{1999JCP.154.284P}\footnote{The code has been well tested and, despite the controversy, is widely used in numerical AMR--based MHD simulations.}.  The $\bmath{\nabla \cdot B}$ terms are left in all of our equations to emphasize this fact: \emph{$\bmath{\nabla\cdot B}$ is \emph{not} explicitly zero in {\sc FLASH}, but constrained below truncation--level error after each timestep} \citep{1999JCP.154.284P}.  Considering these terms during a timestep is thus important in guaranteeing the stability of any computational scheme applied to {\sc FLASH}'s MHD module.  We find it helps decrease $\bmath{\nabla \cdot B}$ by a couple orders of magnitude in longer non--steady computational runs, however it may not be important in more steady flows.  Note that we are also using a method employed in {\sc FLASH} which eliminates $\bmath{\nabla \cdot B}$ as it's created through a simple diffusive method (this is step is performed after each timestep). 

The evolution of $\bmath{\nabla \cdot B}$ in our isothermal C--shock test (\S\ref{sec:codetest}) evolves very much the same with or without these extra terms (there is only a minor advantage of having such terms in the code).  As the shock reaches a steady state, the average value remains constant.  We note that this former case involves small integration times of a few hours and a steady final state. Our simulations of the pre--stellar collapse of a Bonnor--Ebert sphere performed in a companion paper is continually evolving over integration times of a few days.  It shows orders of magnitude improvement when these terms are added and more than 10 orders of magnitude improvement to similar runs performed by the SPH simulations of \citet{HW2004a}.  In a general sense it may be advisable to only use these terms if $\bmath{\nabla \cdot B}$ becomes an issue (e.g. non--steady states, long integration times).  Otherwise, skipping the lines of code pertaining to these terms could save important computing time.

%%%%%%%%%%%%%%%%%%%%%%%%%%%%%%%%%%%%%%%%%%%%%%%%%%%%%%%%%%%%%%%%%%%%%%%%%%%%%%%%
%%%%%%%%%%%%%%%%%%%%%%%%%%%%%%%%%%%%%%%%%%%%%%%%%%%%%%%%%%%%%%%%%%%%%%%%%%%%%%%%
%%%%%%%%%%%%%%%%%%%%%    Testing the Code    %%%%%%%%%%%%%%%%%%%%%%%%%%%%%%%%%%
%%%%%%%%%%%%%%%%%%%%%%%%%%%%%%%%%%%%%%%%%%%%%%%%%%%%%%%%%%%%%%%%%%%%%%%%%%%%%%%%
%%%%%%%%%%%%%%%%%%%%%%%%%%%%%%%%%%%%%%%%%%%%%%%%%%%%%%%%%%%%%%%%%%%%%%%%%%%%%%%%
\section{TEST CASES -- OBLIQUE C--SHOCKS}\label{sec:codetest}

To test the code we have performed a detailed check  with an isothermal C--shock in addition to a second C--shock test which we believe is the first quantitative test in practice of an ambipolar diffusion code with a full non--isothermal energy equation.  Our hope is that this latter test is applied to future codes developing ambipolar diffusion with proper energy considerations.  

%%%%%%%%%%%%%%%%%%%%%%%%%%%%%%%%%%%%%%%%%%%%%%%%%%%%%%%%%%%%%%%%%%%%%%%%%%%%%%%%
%%%%%%%%%%%%%%%%%%%%%%%%%%%%%%%%%%%%%%%%%%%%%%%%%%%%%%%%%%%%%%%%%%%%%%%%%%%%%%%%
%%%%%%%%%%%%%%%%%%%%%%%%%%%%%%    C--shocks    %%%%%%%%%%%%%%%%%%%%%%%%%%%%%%%%%%
%%%%%%%%%%%%%%%%%%%%%%%%%%%%%%%%%%%%%%%%%%%%%%%%%%%%%%%%%%%%%%%%%%%%%%%%%%%%%%%%
%%%%%%%%%%%%%%%%%%%%%%%%%%%%%%%%%%%%%%%%%%%%%%%%%%%%%%%%%%%%%%%%%%%%%%%%%%%%%%%%
C--shocks are continuous transitions of unconserved HD variables mediated by elastic collisions between charged and neutral particles \citep{1993ARA&A..31..373D}.  Without ambipolar diffusion (e.g.~ideal coupling) such flow transitions are highly discontinuous.  C--shocks have quickly become a common test case for ambipolar diffusion codes due to recent interest that has sprung up in a variety of subjects, particularly in turbulence models \citep{1995ApJ...442..726M,1997A&A...326..801S,1997ApJ...491..596M,2003MNRAS.344.1210F,2006MNRAS.366.1329O,2006ApJ...653.1280L}.  Tests for ambipolar diffusion, with a given analytic solution, are very hard to generate due to the complex nature of the equations.  The C--shock has provided a very convenient test case for the papers mentioned, all of which are isothermal.  As astrophysical problems generally involve shocks, we will eventually require full energy considerations with heating and cooling.   A test for such simulations which include the effects of drift heating and energy diffusion does not exist to our knowledge.  However, further development of the idea of C--shocks can be expanded to include these effects.  We present the first such test of a non--isothermal ambipolar diffusion energy code that we are aware of in the literature, in the hopes it will be used as a standard test in the future.  Alongside this we perform an isothermal C--shock test to independently test the induction equation.

\subsection{Basic setup}\label{sec:cshocksetup}
\begin{figure*}
\includegraphics[width=84mm]{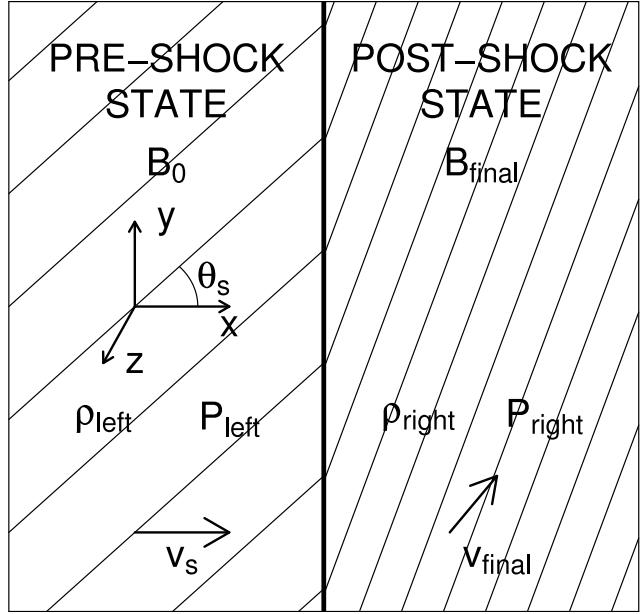}
\caption{\label{fig:test2}
Our initial setup of the oblique C--shock.  The left, pre--shock state is $4\ L_\mathrm{shock}$ in length while the right, post--shock state is $8\ L_\mathrm{shock}$ in length.  The y and z lengths are both (1--2) $\ L_\mathrm{shock}$ in length, for high and low resolution runs respectively.  After a while the shock will smooth out the discontinuity and settle to form a static C--shock.}
\end{figure*}

In Fig.~\ref{fig:test2} the basic idea of an oblique C--shock is illustrated in our initial setup, such that the field $\bmath{B}$ is initially oblique to the normal of the shock front by an angle $\theta_s$.  The discontinuity will be transformed into a more continuous transition as the neutrals sift through the ions and feel the magnetic forces through collisions (modeled as friction).  A final steady and continuous transition is obtained.  The dimensionality of the problem is two, so we run the simulation in permutated orientations of x, y and z to fully test our code.  Nonetheless, each run involves a 3D tube.  The boundary conditions (for the standard permutation presented here) are inflow at the left x  boundary and outflow at the right x boundary, while periodic conditions are used for every other boundary.  For 1D plots we take data from a central row of cells in our box (though no particular row appears different).  

The C--shock analytical solutions are based on work from the 80's and 90's \citep{1980ApJ...241.1021D,1986MNRAS.220..133D,1987ApJ...321..321W,1990MNRAS.246...98W,1991MNRAS.250..523W,1991MNRAS.251..119W}, where the first paper deals with the oblique shocks we present here.  The equations presented therein offer a working analytical solution to a steady state, isothermal, two--fluid C--shock, mediated by ambipolar diffusion.  

In respect to the equations we present below, we have found some inconsistencies with how the energy equation was reduced in the past to give a differential equation for the pressure.  Diffusive contributions to the MHD energy equation were neglected in \citep{1986MNRAS.220..133D}, although frictional heating can be added as in \citet{1991MNRAS.251..119W} (in addition to cooling).  Note however, \citet{1987ApJ...321..321W} argue that the electromagnetic field cannot transport energy as $\bmath{E\times H}=0$. With this in mind, they find the magnetic terms in the energy equation (save for drift heating) will cancel out when the single fluid approximation is carried through and the shock conditions are applied (in the $\bmath{E}=0$ frame).  We show below that this is not the case when diffusive terms are considered and that magnetic terms indeed play important roles in the energy equation.  The agreement of our test results with our theoretical predictions and the significant difference between our predicted pressure distributions with that of \citet{1991MNRAS.251..119W} certifies our viewpoint.

Also note that since we use a single fluid code, we must make a choice on how to treat the ion density evolution.  A simple choice would be to keep it constant ($\rho_i = \rho_{i_0}$).  To account for this involves adjusting a term in the analytical solution, and we outline how this is done below (as well as how to change the equations to ion mass--conserving form or something else).    

We take the initial conditions of \citet{1995ApJ...442..726M} as their shocks show long transitions of $~ (10-20)\ L_\mathrm{shock}$, indicating strong ambipolar diffusion.  Running these through the analytic solutions we find the values in the post--shock region.  Our initial setup consists of a left state with these pre--shock values and a right state with the post--shock values (Fig.~\ref{fig:test2} and Table \ref{tab:test2}).  We evolve the shock tube and allow it to settle to its final configuration, which we compare to analytical predictions.

\begin{table*}
\caption{\label{tab:test2}
The initial conditions for the oblique C--shock test.  The value of $\gamma_\mathrm{AD}$ is 1.0, and the ion density is kept constant at $\rho_i = 10^{-5}\ \mathrm{cm}^{-3}$.  The temperature is adjusted to 10 K by setting the molecular mass appropriately, given that the sound speed is $c_s = 0.1\ \mathrm{cm\ s}^{-1}$.  The initial field deflection is $\theta_s = \pi/4$ and $B_x = B_0 \cos{(\theta_s)}= \sqrt{4\pi}\cos{(\theta_s)}$ is a constant. Non--isothermal post--shock states are only slightly different than isothermal ones (save pressure).}
\center
\begin{tabular}{|l|r|r|r|r|r|}
\hline
state & $\rho_n\ [\mathrm{g\ cm}^{-3}]$ & $P_n\ [\mathrm{dyne\ cm}^{-2}]$ & $v_x\ [\mathrm{cm\ s}^{-1}]$ & $v_y\ [\mathrm{cm\ s}^{-1}]$ & $B_y\ [\mathrm{G}]$ \\
\hline
\hline
left                       & $1.000$ & $0.0100$ & $v_s=5.000$ & $0.000$ &  $2.507$ \\ 
right$_\mathrm{iso}$       & $8.045$ & $0.0804$ & $    0.621$ & $0.840$ & $23.553$ \\
right$_\mathrm{non-iso}$   & $7.976$ & $0.5$ & $    0.627$ & $0.830$ & $23.313$ \\
\hline
\end{tabular}
\end{table*}

The analytic solutions we derived, found by solving the steady multi--fluid equations for the final shock, consist of a coupled set of two first order ordinary differential equations in parameters $p$ and $b$.  The equations are, where we retain the parameter conventions of \citet{1991MNRAS.251..119W}:

\begin{equation}\label{eqn:c-p}
\begin{split}
& \left(\frac{1-\gamma r_n p}{(\gamma -1)r_n}\right)\frac{d p}{d z} =\\ &  \frac{\gamma_\mathrm{AD}\rho_{i_0}}{v_s}\left[\left(\frac{1}{r_n} + \frac{\gamma}{\gamma-1}p -\frac{s_n+\sin{\theta_s}}{b}\right)r +  \frac{\left(G_n-\Lambda_n\right)}{\gamma_\mathrm{AD}\rho_{i_0}\rho_{n_0} v_s^2}\right]
\end{split}
\end{equation}

\begin{equation}\label{eqn:c-b}
\frac{d b}{d z} = \frac{\gamma_\mathrm{AD}\rho_{i_0}}{v_s} A^2 \left(\frac{r}{b}\right),
\end{equation}

where,

\begin{equation}
r_n = \frac{1}{1-\left(p-p_0\right) - \left(\frac{b^2-b^2_0}{2 A^2}\right)}
\end{equation}

\begin{equation}\label{eqn:c-r}
r = \left(1 - \frac{r_n}{r_i}\right)
\end{equation}

\begin{equation}
r_i = r_n\left(\frac{b^2+\cos^2{\theta_s}}{b r_n\left(s_n+\sin{\theta_s}\right)+\cos{\theta_s}^2}\right)
\end{equation}

\begin{equation}
s_n = \frac{b-b_0}{A^2}\cos^2{\theta_s},
\end{equation}
and $\gamma$ is the adiabatic index of the neutral gas, $\gamma_\mathrm{AD}$ is the collisional constant between ions and neutrals described in \S\ref{sec:adtheory}, $\rho_{n_0}$ and $\rho_{i_0}$ are the initial density states of the neutrals and ions respectively, $v_s$ is the initial speed of the gas towards the shock, $A = v_s / v_A$ is the Alfv\'en number of the pre--shock state and $v_A = B_0 / \sqrt{4\pi\rho_{n_0}}$ is the Alfv\'en velocity.  The velocity compression of the neutral and ionized gases are described by the parameters $r_i = v_s/v_{i_x}$ and $r_n = v_s/v_{n_x}$.  The dimensionless pressure is $p = P_n / \left(\rho_{n_0} v_s^2\right)$, and the dimensionless field (in the y--direction) is $b = B_y / B_0$, where $B_0$ is the total initial field magnitude.  The deflection of the field is quantified by $s_n = \left(v_{n_y} B_x\right)/\left(v_s B_0\right)$, as the deflected field will accelerate a velocity in the y direction as it moves through the gas.  The x--component of the field stays constant due to the symmetry in the problem, and in our runs the ionization is kept constant (though ion velocity shocks strongly).  The parameter $r$ defines what type of chemistry is used or if ion mass is simply conserved.  The value given in Equation (\ref{eqn:c-r}) is for a constant ion density. For ion mass conservation (as in \citet{1991MNRAS.251..119W}) the value is $r=(r_i-r_n)$.  For a general chemistry, $r = f(x)\left(1-r_n/r_i\right)$ (such that one can derive $\rho_i = f(x)\rho_{i_0}$, where $f(x)$ is some function in the x direction which may include variables such as $r_n(x)$). 

Drift heating is governed by the parameter:
\begin{equation}
\begin{split}
G_n = &\ \gamma_\mathrm{AD}\rho_{i}\rho_{n}\parallel\bmath{v}_n -\bmath{v}_i\parallel^2 = \mu_{0}\beta_\mathrm{AD}\parallel \bmath{J \times B} {\parallel}^{2}\\
 = &\ \gamma_\mathrm{AD}\rho_{n_0}\rho_{i_0} v_s^2\frac{r^2}{b^2 r_n}\left(b^2+\cos{\theta_s}^2\right), 
\end{split}
\end{equation}
and $\Lambda_n$ is the cooling rate.  The analytical solutions prove to be rather difficult to solve without cooling (via a simple Runge--Kutta 4 technique).  By including a cooling rate as done in  \citet{1991MNRAS.251..119W}, solutions can be obtained that are significantly different than un--heated shocks yet realistically integratable. We take a simplified cooling rate that approximates those of \citet{1983ApJ...270..578L}, relevant to molecular clouds (though meant purely to simplify the computation of this analytic model):
\begin{equation}\label{eqn:cooling}
 \Lambda_n = \left\{
              \begin{array}{ll}
                   \Lambda_{n_0} \left(\frac{\gamma_\mathrm{AD} \rho_{i_0}}{v_s}\right) \left(\frac{p}{p_0}\right)^3\ [\mathrm{erg\ s}^{-1}] & (p > \epsilon_\mathrm{cool} p_0)\\
                   0 & (\mathrm{otherwise}),
              \end{array}
       \right. 
\end{equation}
where $p_0$ is the pre--shock dimensionless pressure, $\epsilon_\mathrm{cool}$ defines at what pressure the cooling turns on and $\Lambda_{n_0}$ is a fudge factor.  We found  $\Lambda_{n_0}= 5\times 10^{-5}$ and $\epsilon_\mathrm{cool}=50$ provide good analytical graphs at a reasonable computational cost, while encapsulating the shock in a similar lengthscale to the isothermal version of the C--shock.  It was important in our case to keep the shock length small as we ran the shock tests in 3D cartesian coordinates.

The lengthscale of the shock is given as $L_\mathrm{shock} = \sqrt{2} v_A t_\mathrm{flow}$, where $t_\mathrm{flow}=1/(\gamma_\mathrm{AD}\rho_{i_0})$ is the time--scale for the shock.  We use these parameters in our simulation setup, having a box $12\times($1-2$)\times($1-2$)\ L_\mathrm{shock}$ in dimensions (depending on resolution), and printing plot files every $0.1\ t_\mathrm{flow}$ (for a total of 500 files).  The evolution consists of a large outward propagating density enhancement which is ejected from the box as the flow settles to form a C--shock (we consider the C--shock `steady' at about $20\ t_\mathrm{flow}$). Note that while this density wave is in the box, timesteps are smaller by about an order of magnitude.  Because of this, a longer box means longer computation times.   

Note that our solution for the pressure derivative in equation \eqref{eqn:c-p} is significantly different than that derived in \citet{1986MNRAS.220..133D} and used in \citet{1991MNRAS.251..119W}.  The difference is the additional term on the right hand side of Equation (\ref{eqn:c-p}):
\begin{equation}
\frac{\gamma_\mathrm{AD}\rho_{i_0}}{v_s}\left[\left(\frac{1}{r_n} -\frac{s_n+\sin{\theta_s}}{b}\right)r \right].
\end{equation}

\begin{figure*}
\includegraphics[width=84mm]{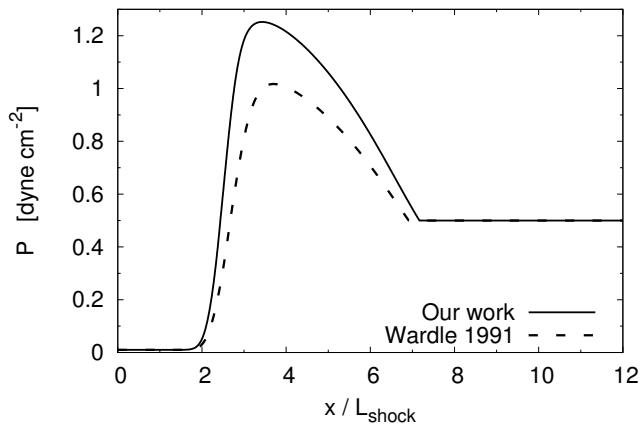}
\caption{\label{fig:pres-comp}
Pressure profiles comparing our analytic solution to that of \protect\citet{1986MNRAS.220..133D} and \protect\citet{1991MNRAS.251..119W}.  A significant difference is apparent.}
\end{figure*}

We compare our solution to that of \citet{1991MNRAS.251..119W} in Fig.~\ref{fig:pres-comp} for an adiabatic index of $\gamma=1.4$ and the same cooling rate defined above and using the values presented in Table \ref{tab:test2}.  The significant difference between the results is quickly apparent.  See typical isothermal values of pressure in Table (\ref{tab:test2}) or errors in Fig.~\ref{pres-300} to understand the scope of the numbers in this figure.  Other quantities have only minor differences. 

\subsection{Numerical results}
\begin{figure*}
\begin{tabular}{cc}
\includegraphics[width=84mm]{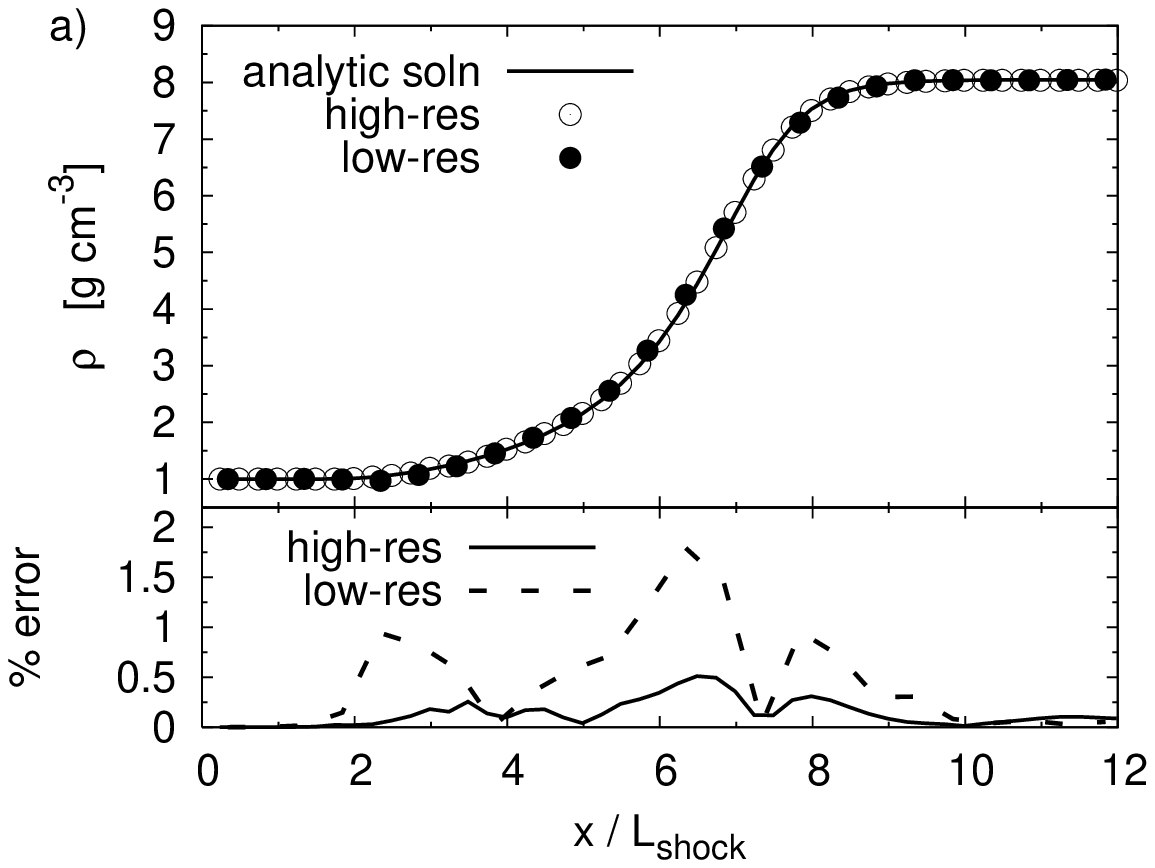} & \includegraphics[width=84mm]{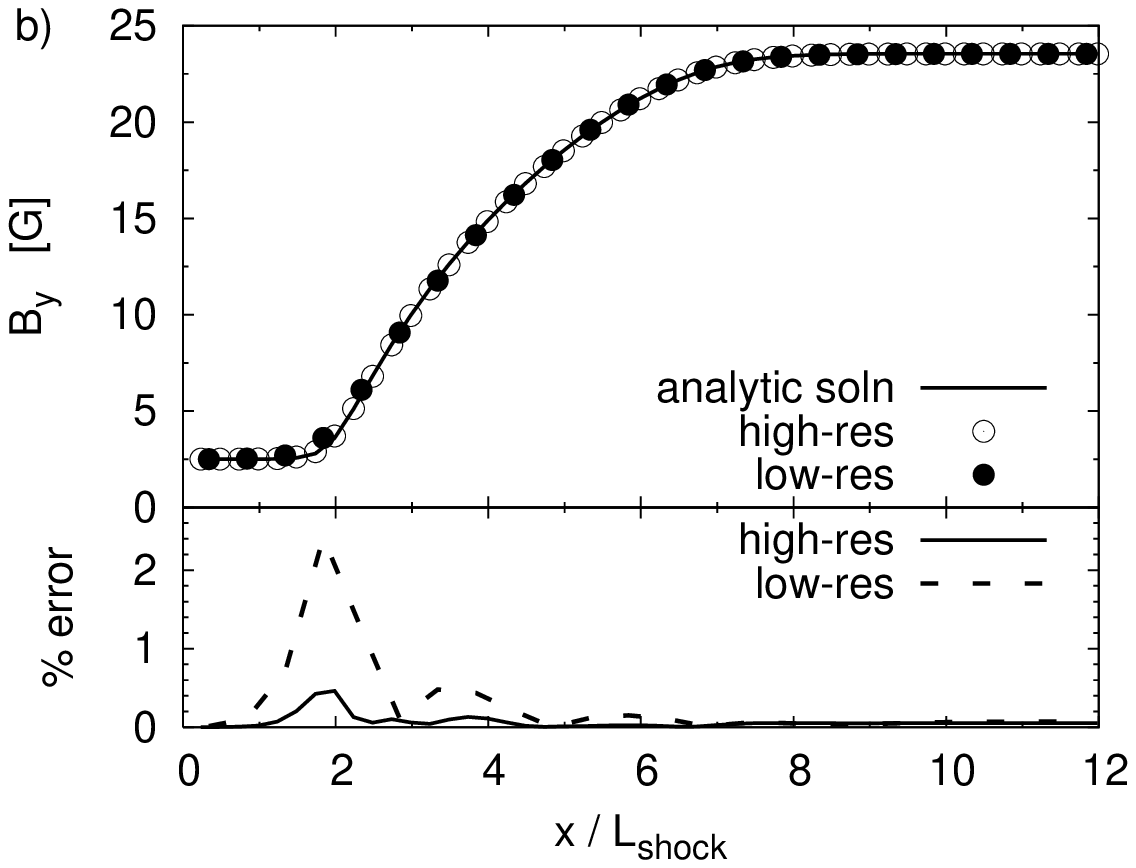}\\
\includegraphics[width=84mm]{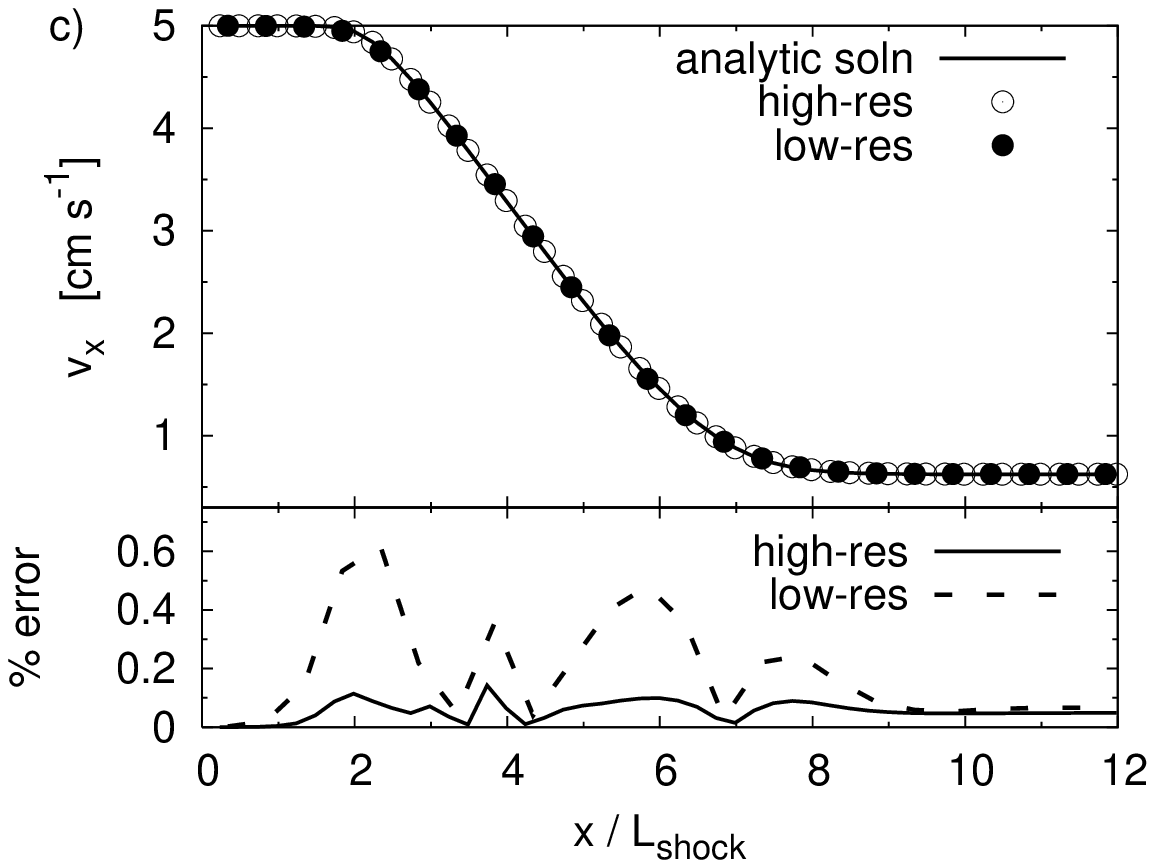} & \includegraphics[width=84mm]{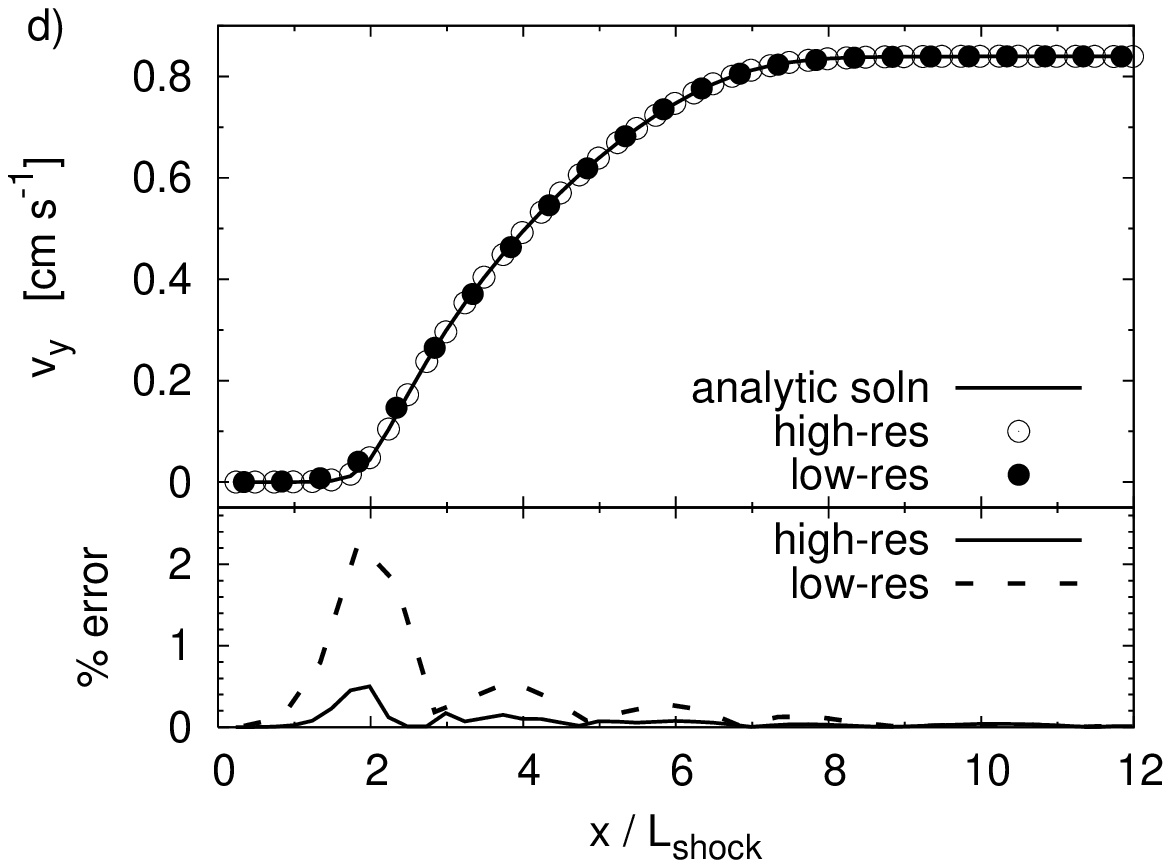}\\
\end{tabular}
\caption{\label{fig:cshock-iso}
Profiles comparing simulations of isothermal C--shocks with analytical solutions at different resolutions for (a) density, (b) y--component of magnetic field, (c) x--component of velocity and (d) y--component of velocity.  Accuracy is high and convergence is evident.  Pressure distributions are identical (though scaled) to density.}
\end{figure*}

In the following subsection we present comparisons of isothermal and  non--isothermal C--shock simulations relative to analytical solutions.  The error is calculated as in \citet{1995ApJ...442..726M}:
\begin{equation}
\% \ \mathrm{error} = 100\left|\frac{q_\mathrm{analytic} - q_\mathrm{numerical}}{\mathrm{max}\left(q_\mathrm{analytic}\right)}\right|.
\end{equation}
We employ rather low resolution in these simulations; our `low' resolution has cell sizes of $(1/2)\ L_\mathrm{shock}$ while our `high' resolution run resolves $(1/4)\ L_\mathrm{shock}$ (though in 3D).  For all our models we plot the percent error for the high and low resolution cases.  It can be seen from the figures that the high resolution error is strongly reduced in relation to the low resolution errors.  This is enough to show the high accuracy of our models (giving similar errors to that of \citet{1995ApJ...442..726M} with half the resolution) and demonstrate a clear convergence, thus confirming our implementation of the induction equation.  Our results are in Fig.~(\ref{fig:cshock-iso}).

\begin{figure*}
\includegraphics[width=84mm]{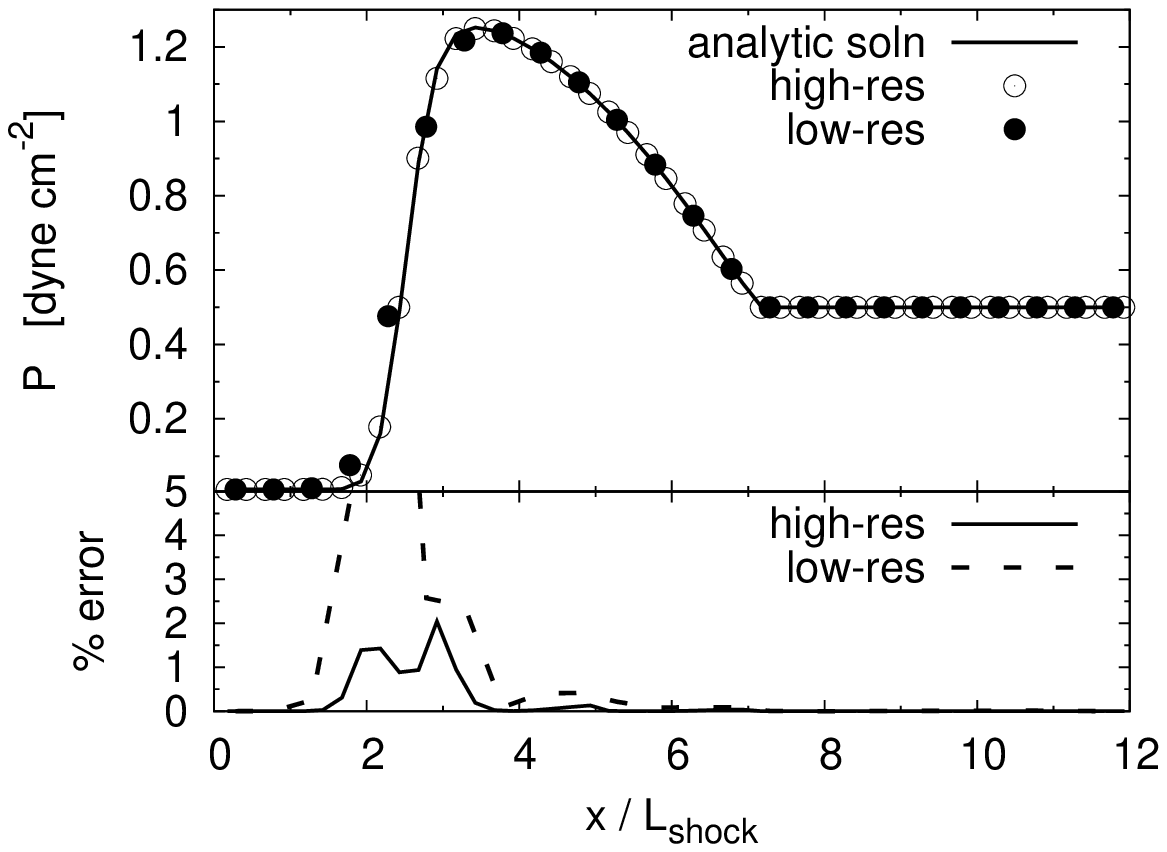}
\caption{\label{pres-300}
Pressure profiles comparing simulations of non--isothermal C--shocks with analytical solutions at different resolutions.  Accuracy is high and convergence is evident.  Temperature distributions are similar (though scaled) to pressure.}
\end{figure*}

\begin{figure*}
\begin{tabular}{cc}
\includegraphics[width=84mm]{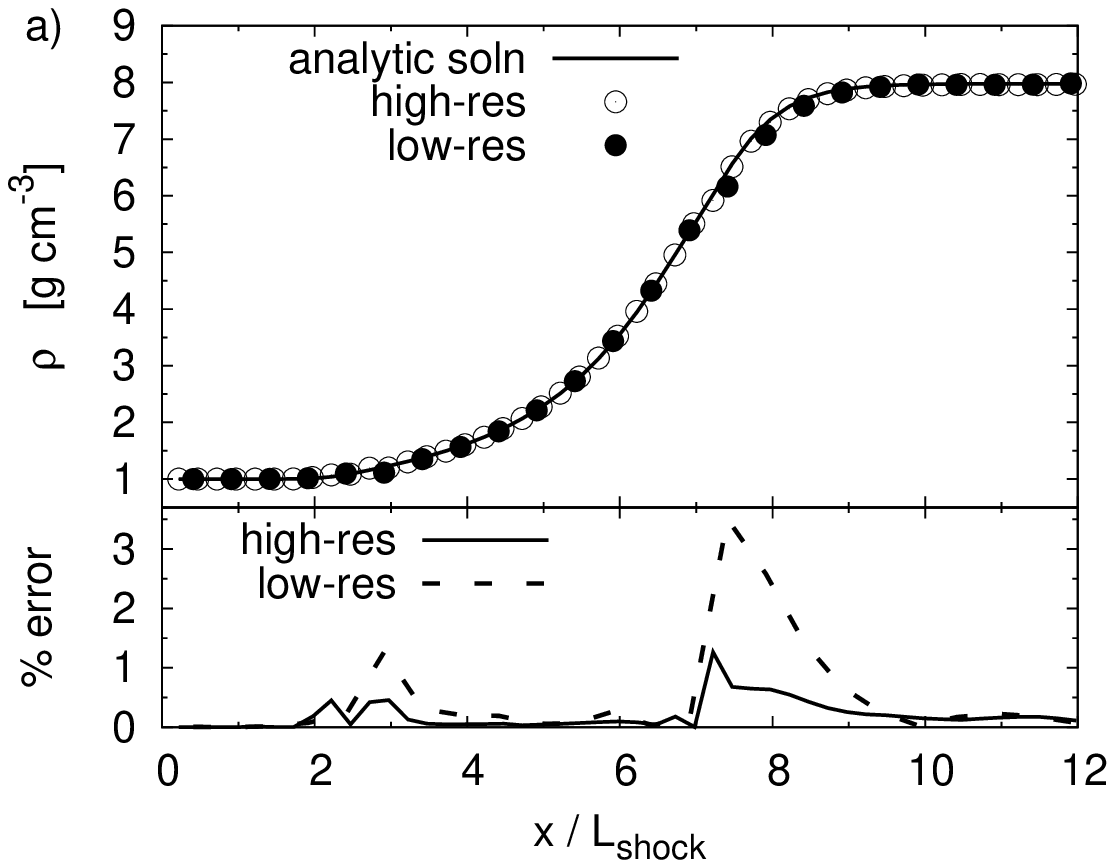} &
\includegraphics[width=84mm]{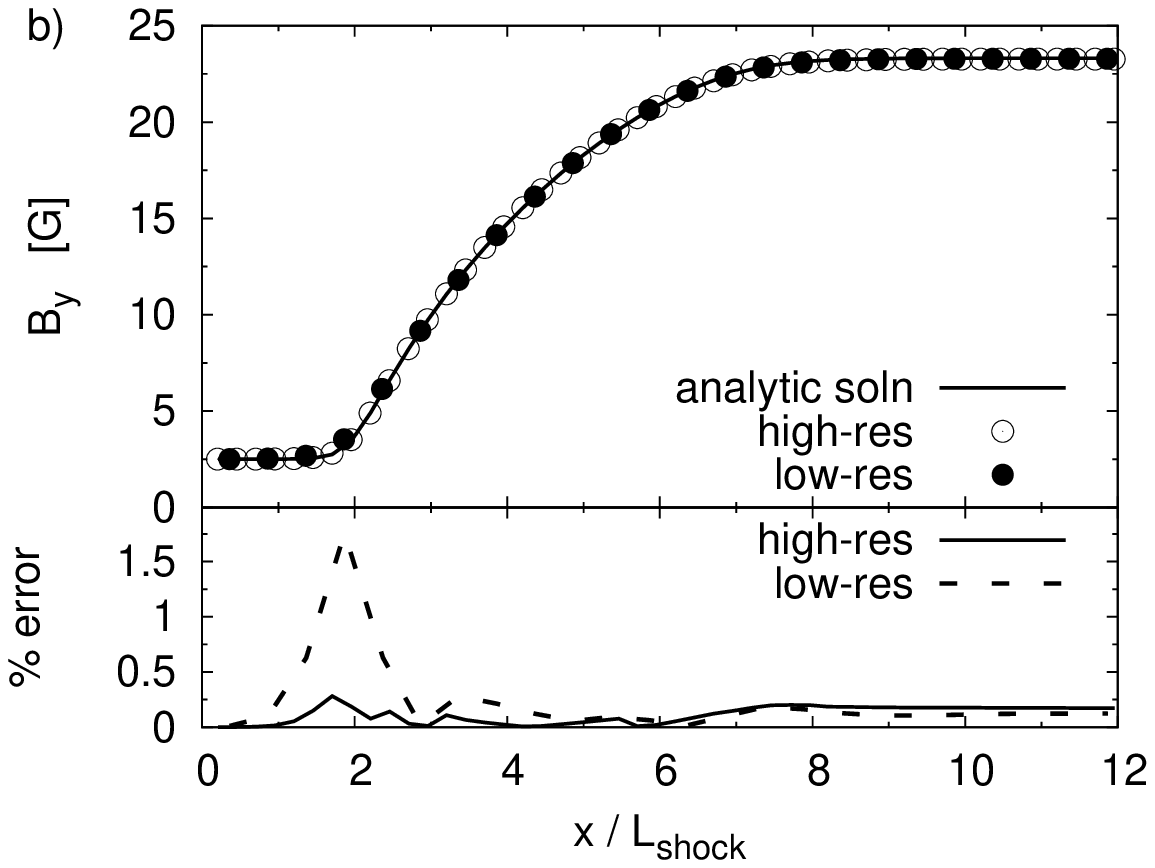}\\
\includegraphics[width=84mm]{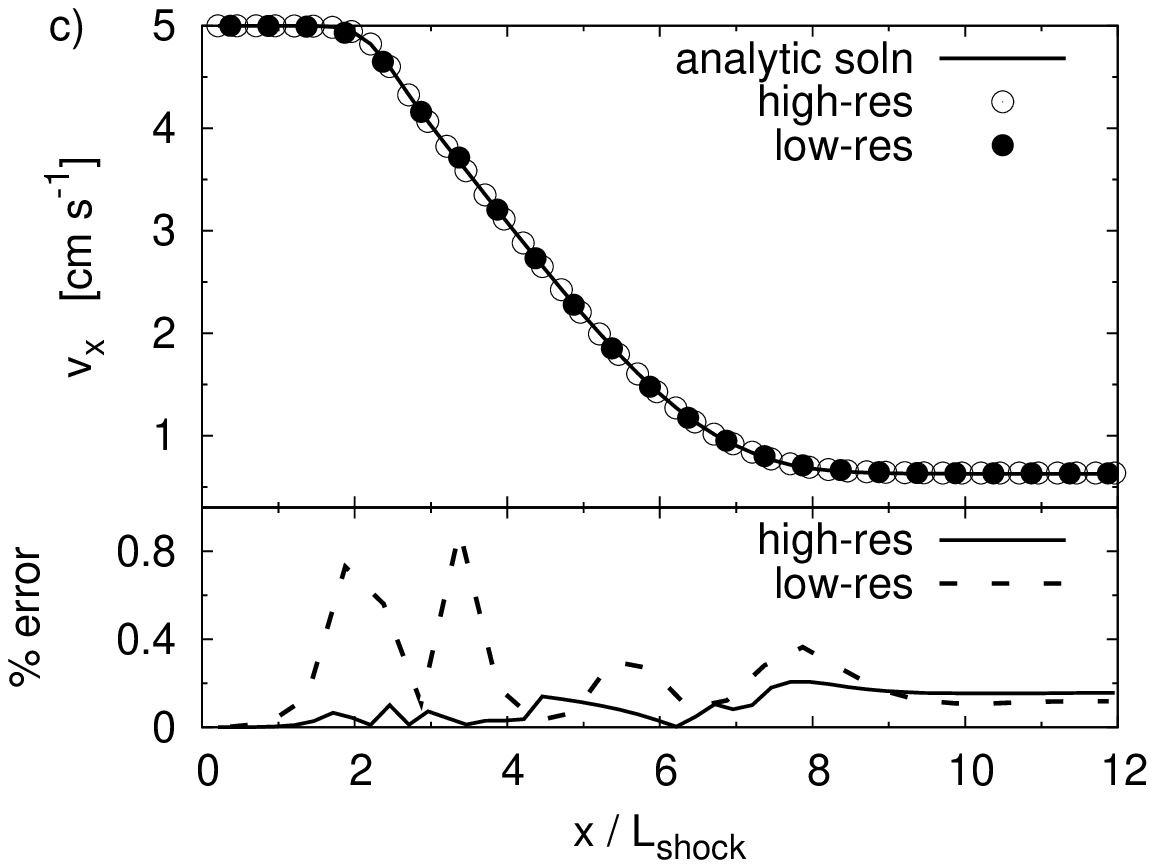} &
\includegraphics[width=84mm]{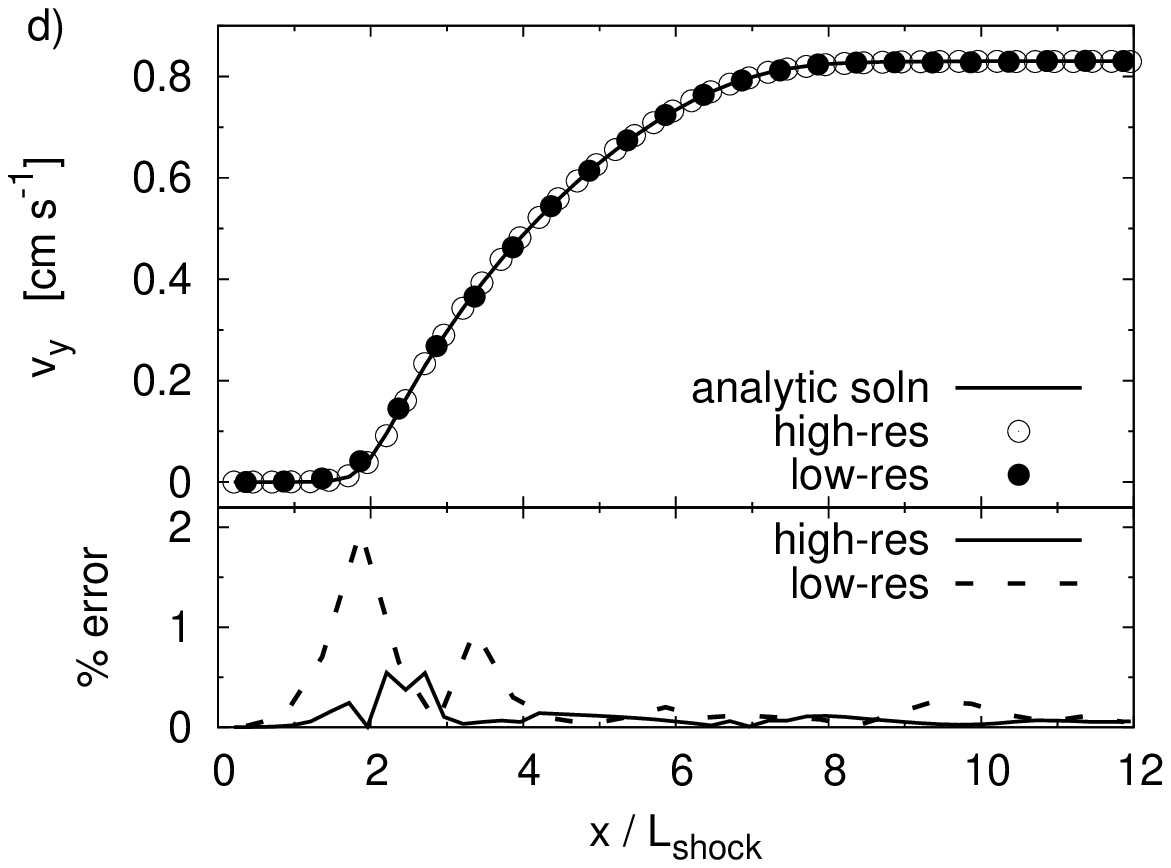}\\
\end{tabular}
\caption{\label{fig:cshock-adi}
Profiles comparing simulations of non--isothermal C--shocks with analytical solutions at different resolutions for (a) density, (b) y--component of magnetic field, (c) x--component of velocity and (d) y--component of velocity.  Accuracy is high and convergence is evident on the left state.}
\end{figure*}

Finally we present the results of the non--isothermal run, testing our energy terms.  For these simulations we use an adiabatic index of $\gamma=1.4$.  These simulations also show high accuracy and a clear convergence.  Our results are seen in Fig.~(\ref{pres-300}) and (\ref{fig:cshock-adi}).  The sharp edge in the post--shock region of Fig.~(\ref{pres-300}) is a result of adjusting the parameters in our toy cooling model (\ref{eqn:cooling}).  This was done to properly encapsulate a stable non--isothermal C--shock on a grid size similar to the isothermal C--shock.

It is clear from the above tests that our code performs admirably in all facets. 

%%%%%%%%%%%%%%%%%%%%%%%%%%%%%%%%%%%%%%%%%%%%%%%%%%%%%%%%%%%%%%%%%%%%%%%%%%%%%%%%
%%%%%%%%%%%%%%%%%%%%%%%%%%%%%%%%%%%%%%%%%%%%%%%%%%%%%%%%%%%%%%%%%%%%%%%%%%%%%%%%
%%%%%%%%%%%%%%%%%%%%%    quasi--static Collapse    %%%%%%%%%%%%%%%%%%%%%%%%%%%%%%
%%%%%%%%%%%%%%%%%%%%%%%%%%%%%%%%%%%%%%%%%%%%%%%%%%%%%%%%%%%%%%%%%%%%%%%%%%%%%%%%
%%%%%%%%%%%%%%%%%%%%%%%%%%%%%%%%%%%%%%%%%%%%%%%%%%%%%%%%%%%%%%%%%%%%%%%%%%%%%%%%
\section{TEST CASE -- QUASI--STATIC COLLAPSE}\label{sec:fm93}

This test involves matching qualitatively the characteristics (such as time--scales and curve features) of the quasi--static collapse of a thermally critical, but otherwise magnetically subcritical, non--rotating core.  In order to study the effects of AD in an axisymmetric geometry, \citet{FM1993b} (\citetalias{FM1993b} hereafter) set up an initial state with a periodic boundary conditions (along all surfaces) meant to represent an infinite chain of fragments.  They used a static, specially designed adaptive grid to follows the evolution.  Its history as use for a test code for ambipolar diffusion is sporadic, appearing in \citet{SMS1997} (an analytical model) and more recently in SPH models by \citet{HW2004a}. 

Our purpose is to investigate a test problem whose collapse will have similarities with that of the well--studied Bonner--Ebert models. There is considerable evidence that these latter states actually exist \citep*{2001Natur.409..159A}.  Nevertheless, in order to make comparisons with the  \citetalias{FM1993b} calculations, we use the physical parameters from their model 1 within our own spherical initial states.  We present these physical parameters in Table \ref{tab:test1}.  Although the model we use is different from \citetalias{FM1993b},  it produces a thin quasi--static disc of similar characteristics.  In this sense it is studying the same problem, the time--scale between thin disc formation and dynamic collapse.  The difference in models is due to the unique way in which the problem was initially setup in \citetalias{FM1993b}, using periodic boundary conditions on a small cylinder -- whereas we employ an adaptive mesh refinement scheme in a 3D Cartesian grid.  In the end our model studies the problem of an isolated cloud, while the cited authors studied closely packed clouds.

\begin{table}
\caption{\label{tab:test1}
The parameters used in the quasi--static collapse test run.  Note that the ionization is the same described in \S\ref{sec:ions}, where $\mu_\mathrm{mol} = 2.33$.  The sound speed then corresponds to a temperature of $10\ \mathrm{K}$.  The box dimensions extend from $-3.24\ \mathrm{pc}$ to $+3.24\ \mathrm{pc}$.}
\center
\begin{tabular}{|r|r|r|r|r|}
\hline
$R_\mathrm{test}\ [\mathrm{pc}]$  & $n_\mathrm{in}\ [\mathrm{cm}^{-3}]$ & $n_\mathrm{out}\ [\mathrm{cm}^{-3}]$ & $b_\mathrm{test}\ [\mathrm{\mu G}]$ & $c_s\ [\mathrm{km\ s}^{-1}]$ \\
\hline
\hline
$0.75$ & $300.0$ & $30.0$ & $30.0$ & $1.889$ \\ 
\hline
\end{tabular}
\end{table}

\begin{figure*}
\includegraphics[width=84mm]{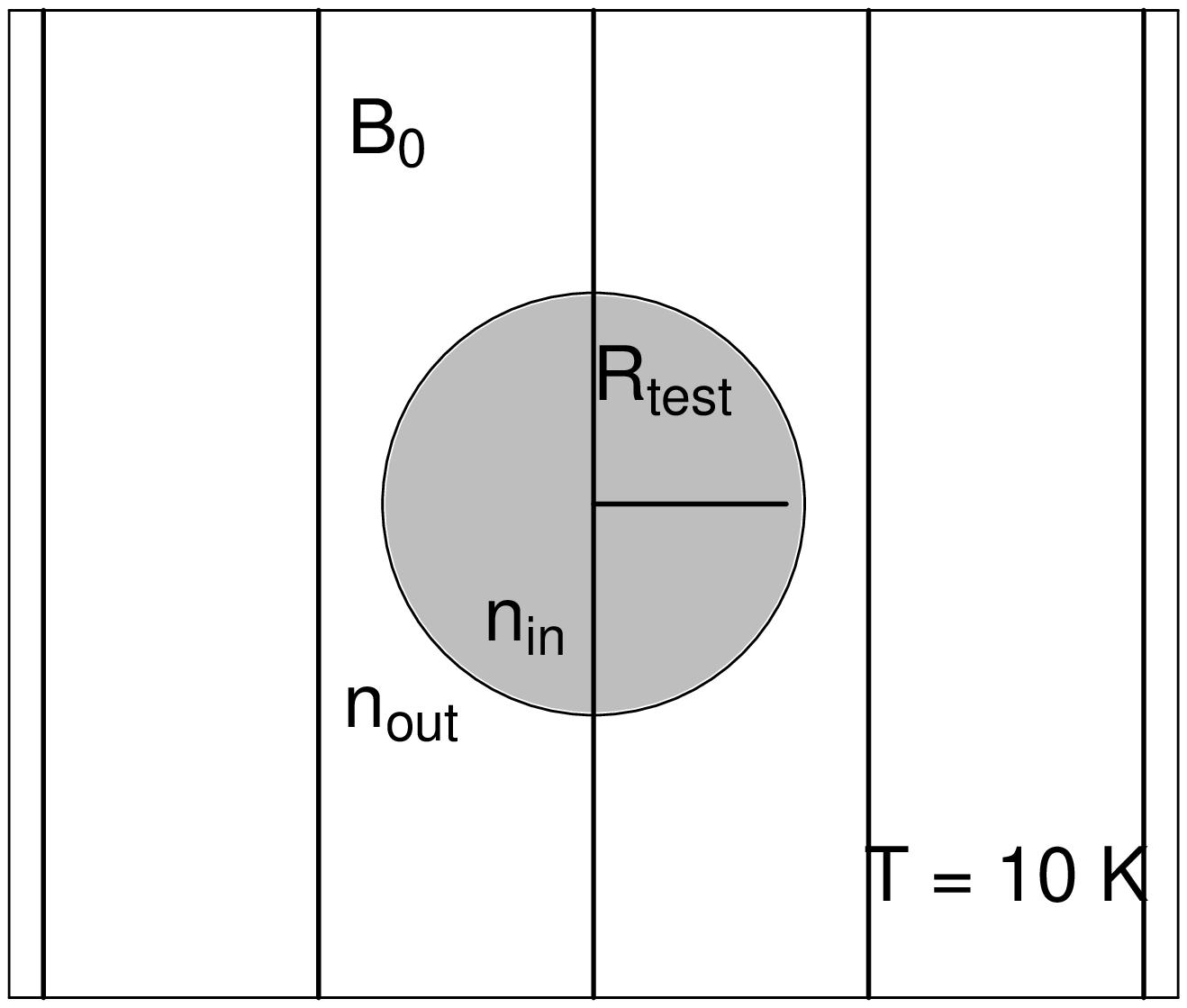}
\caption{\label{fig:test1}
Setup of the the quasi--static collapse test problem.  Our setup of the quasi--static collapse is different than the one presented by \citetalias{FM1993b}.  The key parameters are shown.  The whole box is kept at 10 K throughout the collapse.}
\end{figure*}

Our model consists of a sphere of uniform density $n_\mathrm{in}=300~\mathrm{cm}^{-3}$,  radius $R_\mathrm{test} =0.75~\mathrm{pc}\approx 1.6 \times 10^5~\mathrm{AU}$  , and total mass 30.3 $M_{\mathrm{\sun}}$.  The critical mass for the inner sphere if it were non--magnetized is $M_\mathrm{cr} = 6.6 M_{\mathrm{\sun}}$ \citep{2007arXiv0707.3514M}, so the sphere is thermally unstable.  However, the initial mass--to--flux ratio is $\Gamma = 0.298 <1$, meaning the cloud is supported by its magnetic field.  The sphere is in thermal equilibrium with an external medium of lower density ($\rho_\mathrm{ambient} = 0.1 \rho_\mathrm{core}$).  The setup is shown schematically in Fig.~\ref{fig:test1}.  The model of \citetalias{FM1993b} used a uniform cylinder of density $\rho_\mathrm{core}$, radius $R_\mathrm{test}$ and height above the midplane $Z=R_\mathrm{test}$.  Both models have periodic boundary conditions, thus the latter will have less of an ambient source of material in which to accrete due to it modeling a closely knit cluster of collapsing thin discs (as a consequence of a small box size and periodic boundary conditions).  Also, our initial setup will collapse much quicker into the quasi--static state as it is initially highly unstable.  The setup of \citetalias{FM1993b} is only slightly thermally critical to fragmentation.  We run pure HD and ideal MHD models alongside the non--ideal MHD model to shed light on what is happening physically.

\begin{figure*}
\begin{tabular}{cc}
\includegraphics[width=84mm]{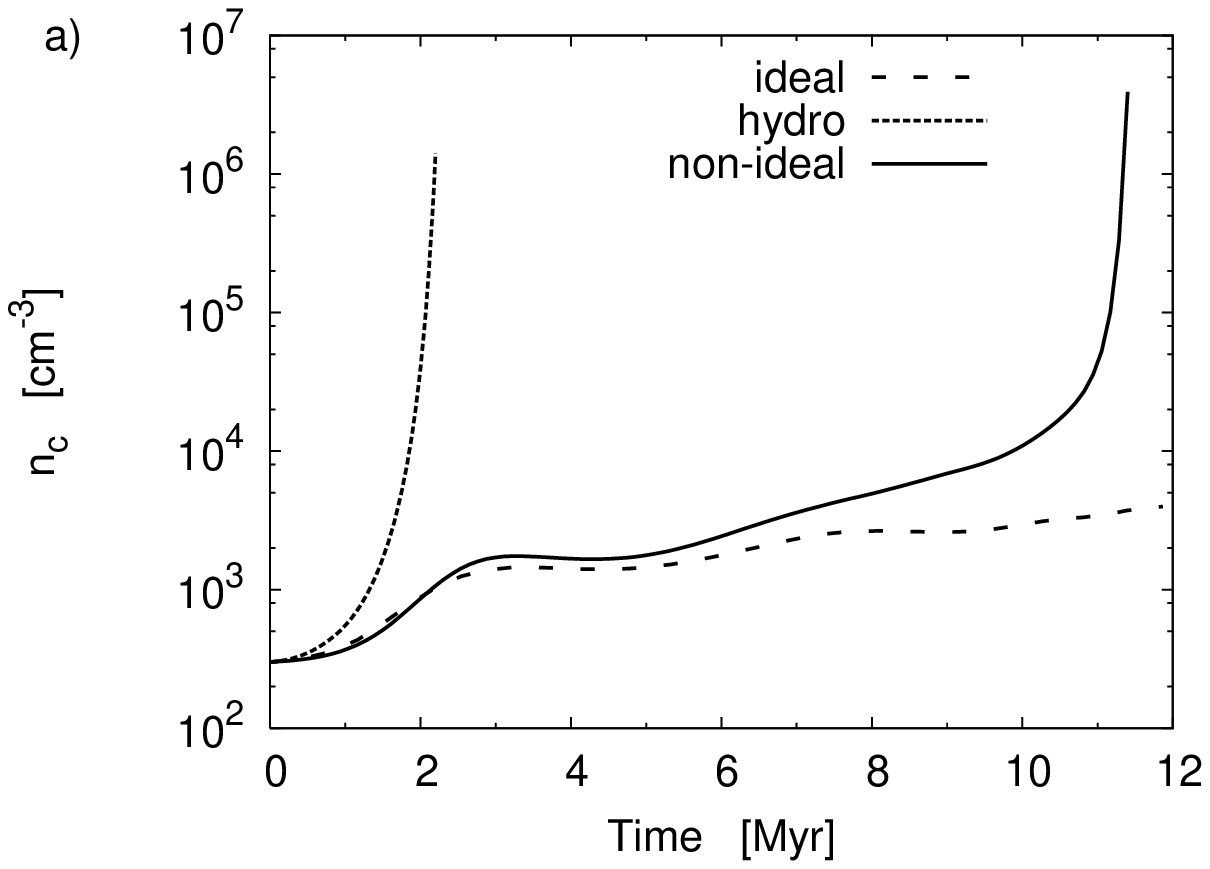} & 
\includegraphics[width=84mm]{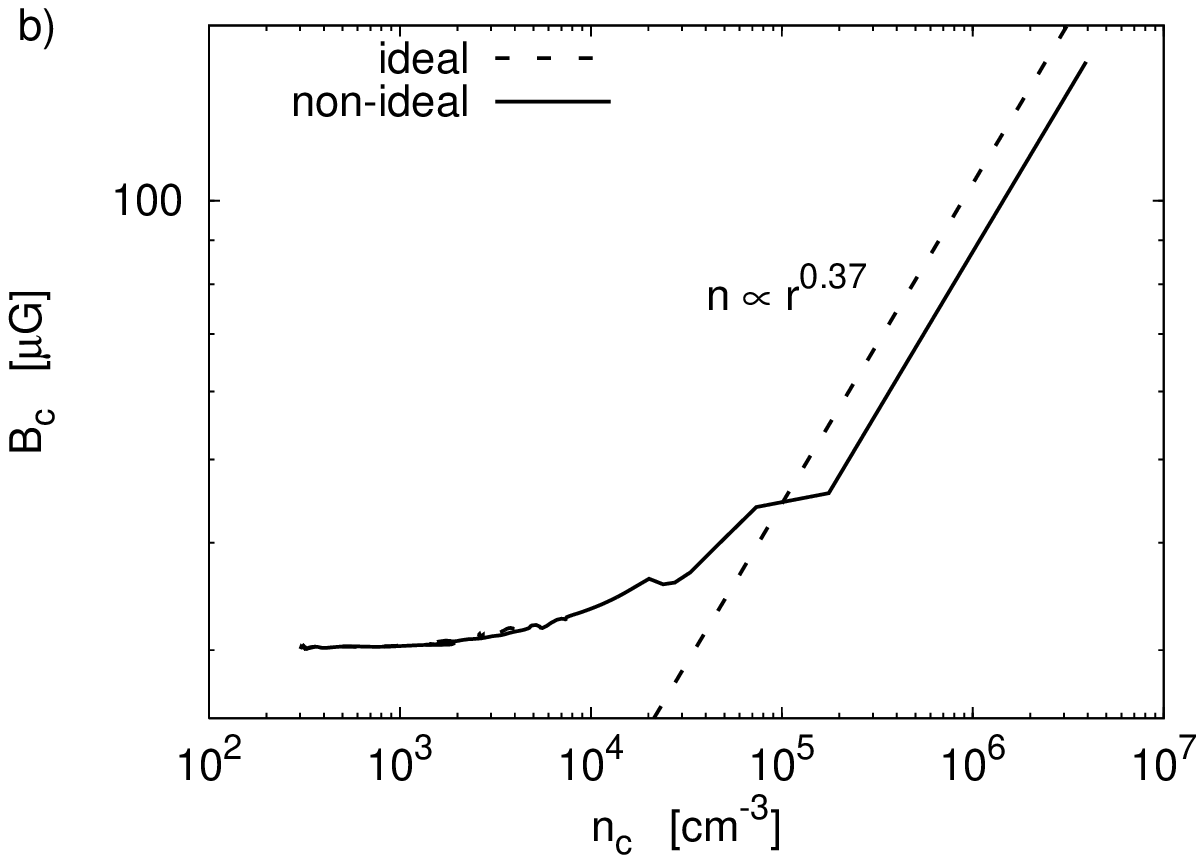}\\
\end{tabular}
\caption{\label{fig:adtest1} 
Results from the first test run showing the evolution density and magnetic field initiated from our quasi--static collapse setup.  We plot (a) central density evolution and (b) the evolution of the central magnetic field.  The ambipolar diffusion run is shown in the solid line and the ideal run is shown in the long dashed line.  A non--magnetized `hydro' run is shown as a short dashed line is a). }
\end{figure*}

The sphere quickly collapses along the field lines into a quasi--static disc (which matches the description of the \emph{thin disc} model described by \citetalias{FM1993b} and many others).  The evolution of the central density is shown in Fig.~\ref{fig:adtest1}.  The initial rise in density is due to the rebounding after the initial infall, as described by \citetalias{FM1993b}.  Gas piles into the disc while magnetic pressure is pushing it out.  At this time (about the free fall time, $t_{\mathrm{ff}} = 1.96$ Myr), without the magnetic field the cloud would have entered a freefall collapse (illustrated by the pure HD curve in Fig.~\ref{fig:adtest1}a).  However, the disc reaches the quasi--stable equilibrium -- where gravitational forces are balanced by magnetic forces.  The infalling gas on larger scales takes a while to respond and  continues to accrete more matter on to the disc.  Magnetic pressure forces then drive out the excess matter and the disc settles into a quasi--equilibrium.  Note that the cited authors reach this characteristic `rebound' peak at 6 Myr, while we reach it at 2 Myr.  

\begin{figure*}
\begin{tabular}{cc}
\includegraphics[width=84mm]{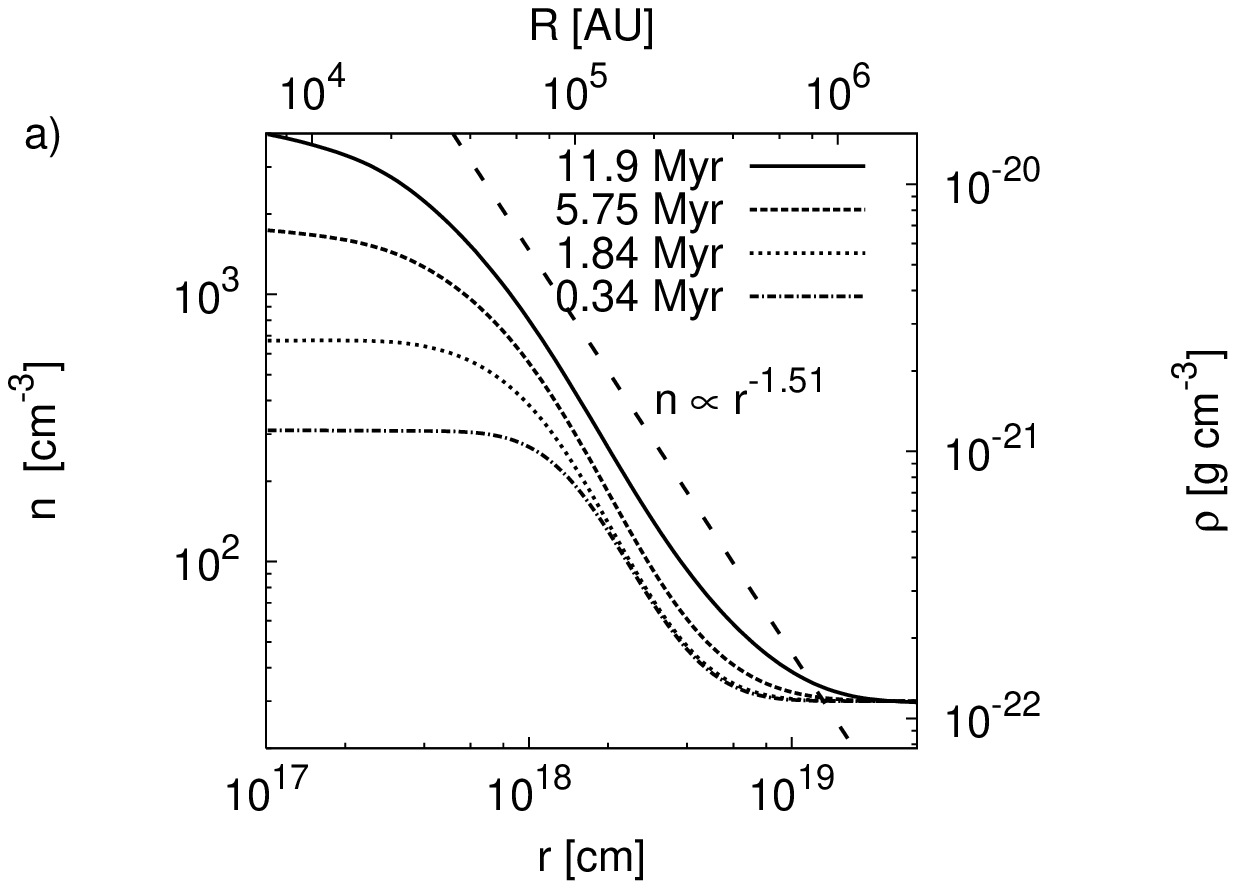} &
\includegraphics[width=84mm]{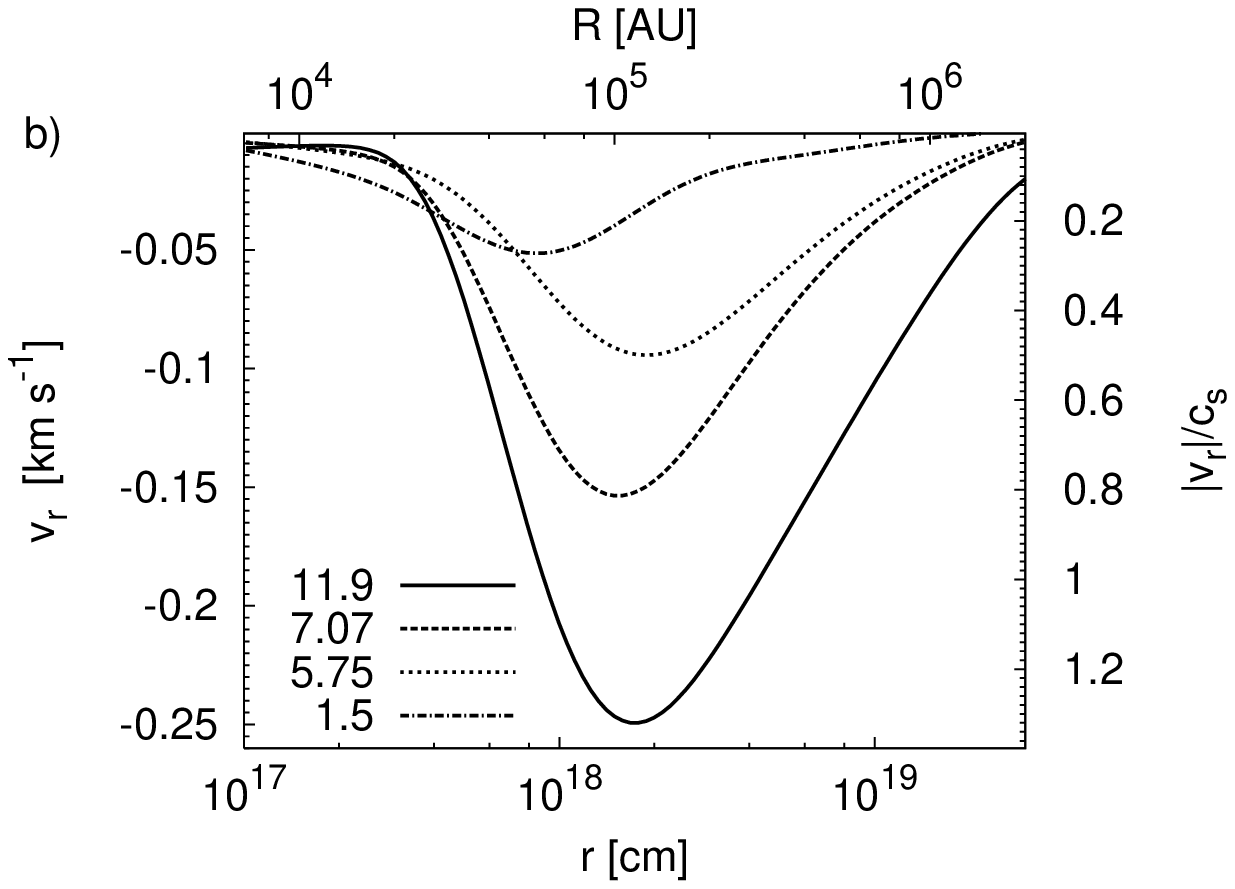} \\
\includegraphics[width=84mm]{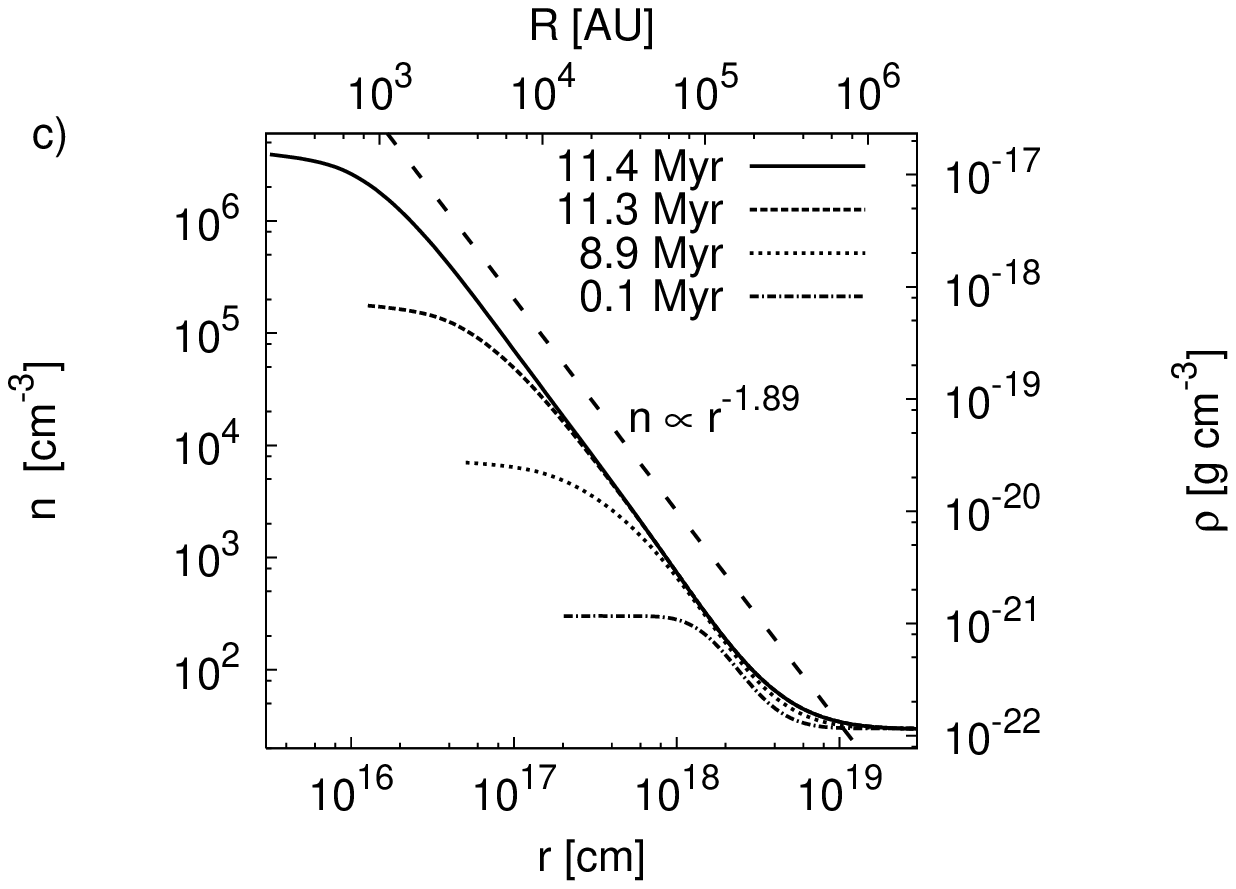} &
\includegraphics[width=84mm]{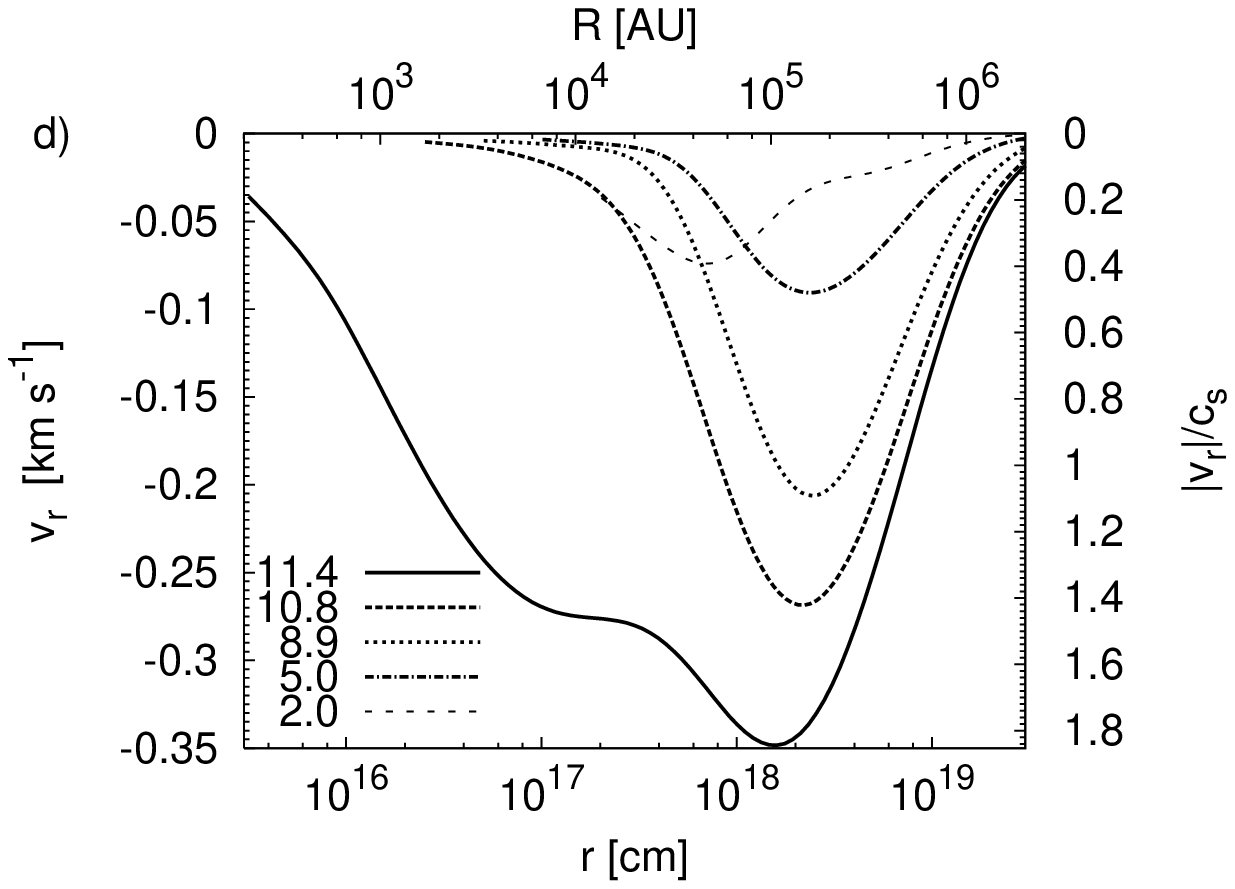} \\
\includegraphics[width=84mm]{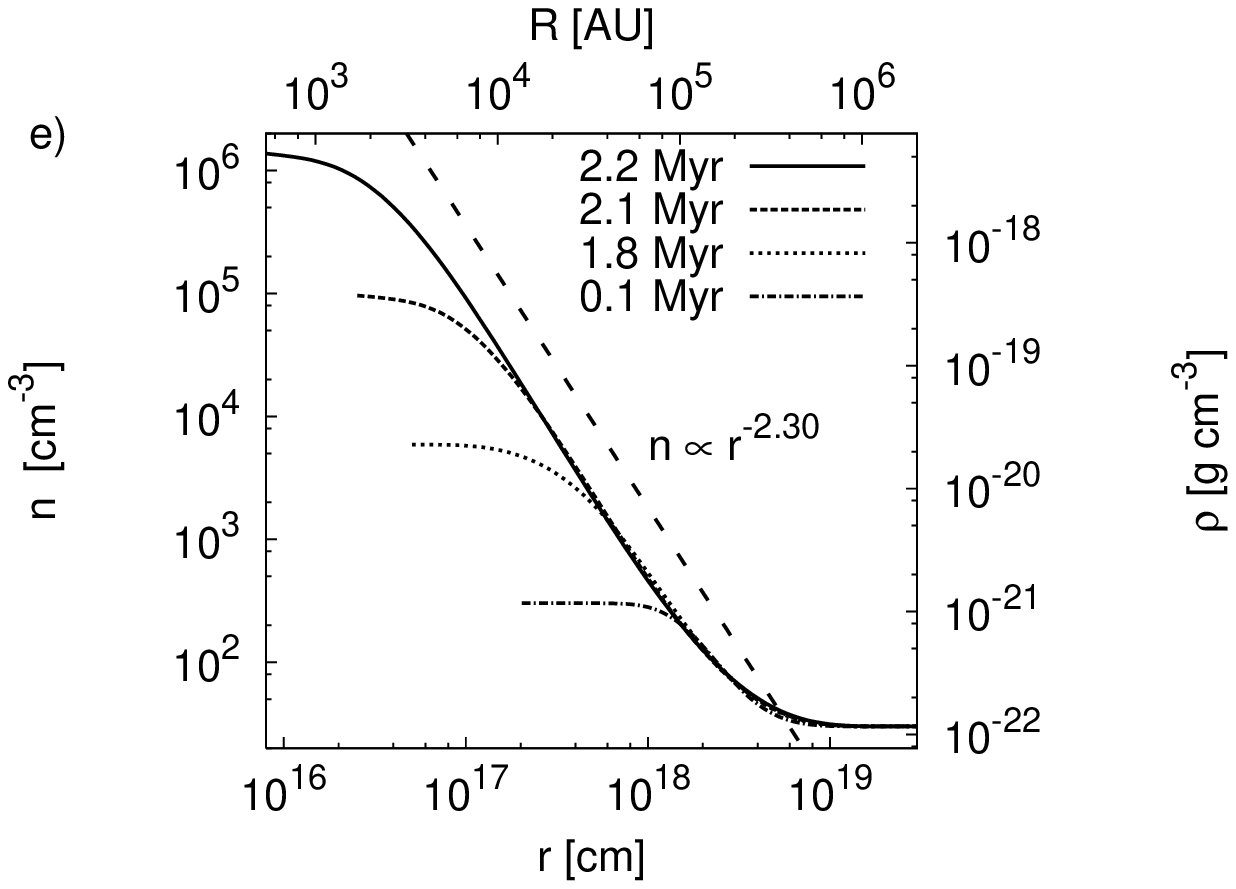}& 
\includegraphics[width=84mm]{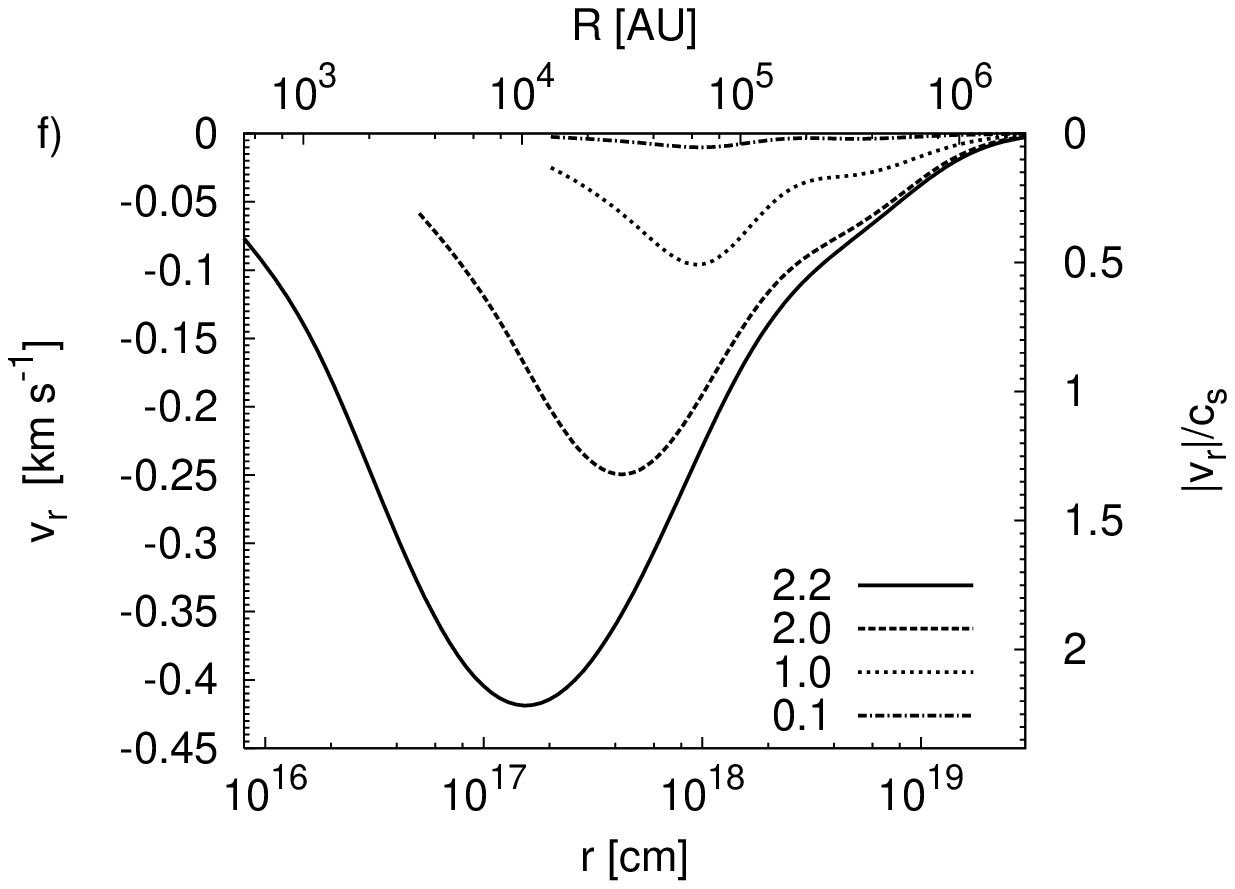} \\
\end{tabular}
\caption{\label{fig:profiles} 
Radial profiles in the quasi--static collapse test at different times for ideal, non--ideal and hydro models.  Profiles are of density and radial velocity in the a-b) ideal, c-d) non--ideal and e-f) hydro runs respectively.  Curves are chosen such that they are evenly spaced.  Values of density and radial velocity are spherically averaged for a given value of r.
}
\end{figure*}

The ideally magnetized case will evolve only marginally after this, accreting material from the ambient medium. This is because the mass to flux ratio of this state is still subcritical (The maximal $\Gamma$ for the ideal model is about $\Gamma = 0.41 < 1$ at the end of the simulation, while it is about 4.90 by the end of the non--ideal run). The non--ideal disc also continues to accrete mass internally as well as from the ambient medium -- its central density evolves at a higher rate than that of the ideal case.  This is illustrated by the evolution of the density as well as the radial velocity as a function of distance, and for different times in the simulations (Fig.~\ref{fig:profiles})\footnote{The profiles are radially binned spherical averages, such that for a quantity $v$, $v(r) =\int v(\bmath{x})d\Omega$, where $d\Omega=d\cos{\theta}d\phi$.}.  Of note is the considerable difference in density ranges for each model.  Inside of 10$^4$ AU the ideal model has very small inflow motions. While, in the later stages of its evolution, the non--ideal model appears more like the purely HD model, both having considerable inflow motions inside of 10$^3$ AU.  This indicates that indeed significant inflow is occurring inside a region of the disc that is supported in the ideal limit.  It is interesting to note how the infall motions resemble the outside--in collapse motions that characterize a Bonnor--Ebert sphere \citep{BP2007}.

The field lines are found to remain straight (until full freefall collapse is reached) despite advanced infall in the AD model.  This  corresponds to neutral gas moving, under the influence of gravity, through the ions attached to the magnetic field.  

Eventually, freefall occurs in the non--ideal model.  This is because ongoing AD in the quasi--equilibrium state gradually increases the mass to flux ratio until a critical state for collapse is achieved, and then surpassed.  The \citetalias{FM1993b} model has a freefall collapse that begins at about 16 Myr while ours begins at about 12 Myr.  The difference is readily understood by observing that each freefall state happens about 10 Myr after the rebounding `bump' occurs and the quasi--static disc is formed.  In this sense our models agree on the time--scale of ambipolar diffusion (although, by having a larger reservoir of material to accrete from, we slightly shortens our time--scale from 10 Myr). 

The radial profiles of Fig.~\ref{fig:profiles}c and Fig.~\ref{fig:profiles}d also agree well with the previous work.  \citetalias{FM1993b} find $d\log{n}/d\log{r}\approx -2$ while we find a value of -1.89.  In areas more supported by the magnetic field, \citetalias{FM1993b} found a profile that goes more like -1.6. In our ideal run, we found a profile of -1.51 (the cited authors do not list detailed properties of the ideal collapse).  It is interesting to note that our HD freefall collapse had a profile of -2.3, indicating a restricted build--up of density in the non--ideal model, clearly due to the magnetic field.  The radial velocity profiles are also similar to \citetalias{FM1993b}, showing similar ranges in $v_r$ and a similar profile.  Supersonic velocities are achieved in all cases.  We do not detect a shock forming in any of our models.   

The evolution of the central magnetic field with density is shown in Fig.~\ref{fig:adtest1}b.  Our curve has the right qualitative features:  a roughly constant field followed by a positive power law.  The power law begins at roughly (2-4) $\times10^{4}$ in agreement with \citetalias{FM1993b}.  At greater densities, there is a distinct power--law relation between these quantities -- we find 
\begin{equation} 
B_c = 0.51 n_c^{\kappa};~\kappa \simeq 0.37
\end{equation}
which compares well, though not exactly, with their result of $\kappa = 0.47$.  We note however that we chose to resolve our runs by every 0.1 Myr and were left with only two data points to calculate $\kappa$.

\begin{figure*}
\begin{tabular}{cc}
\includegraphics[width=84mm]{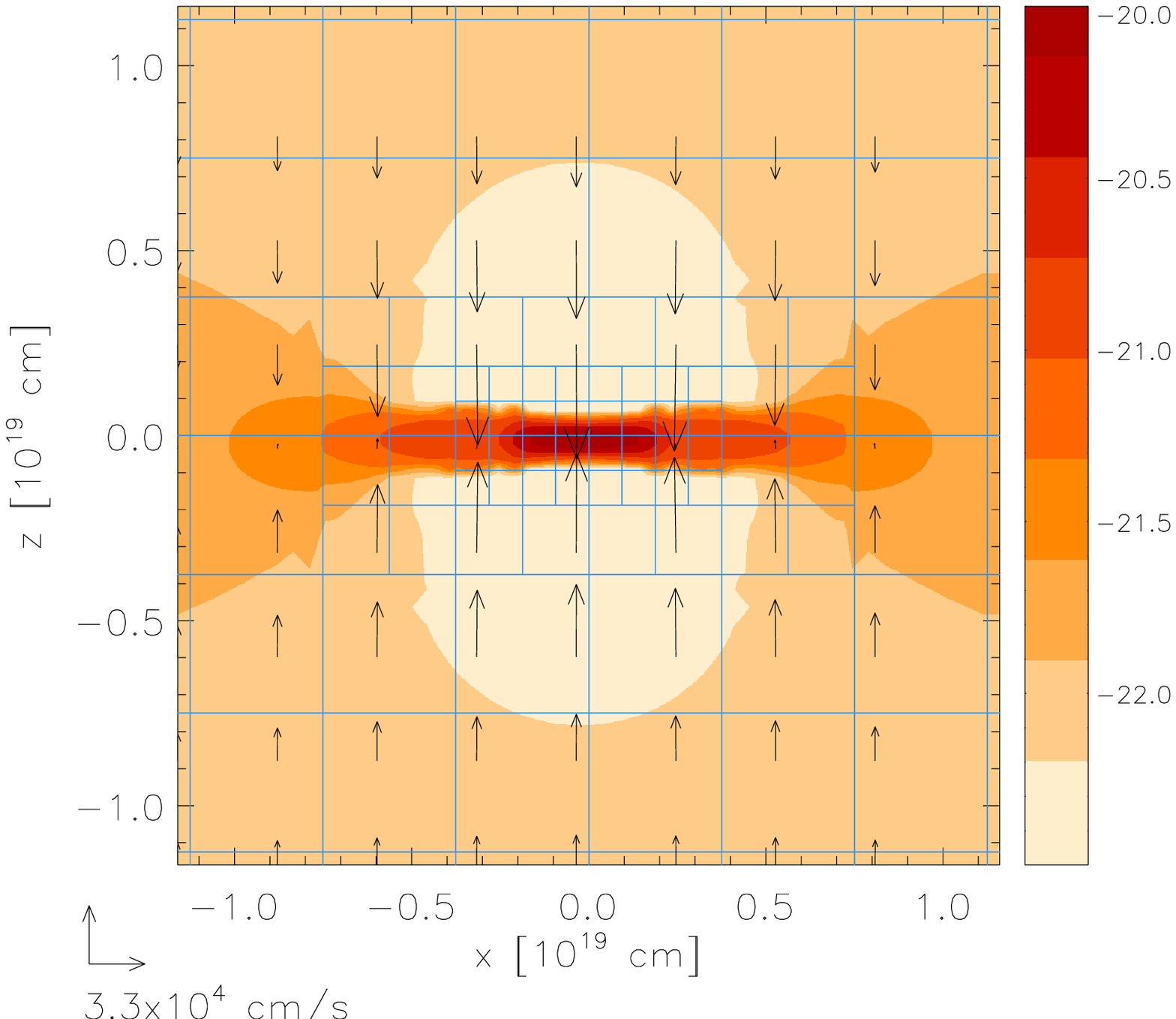} &
\includegraphics[width=84mm]{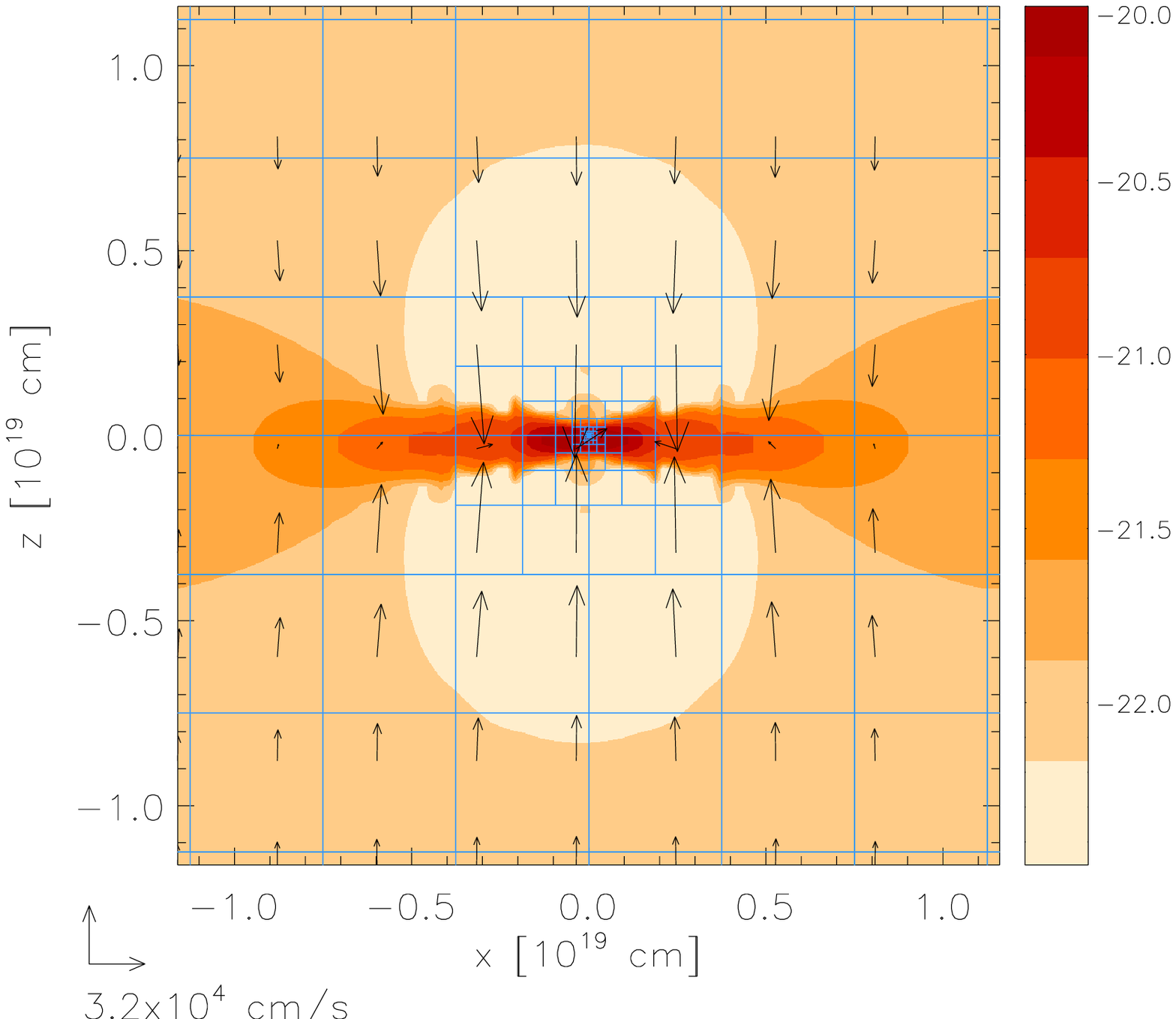} \\
\includegraphics[width=84mm]{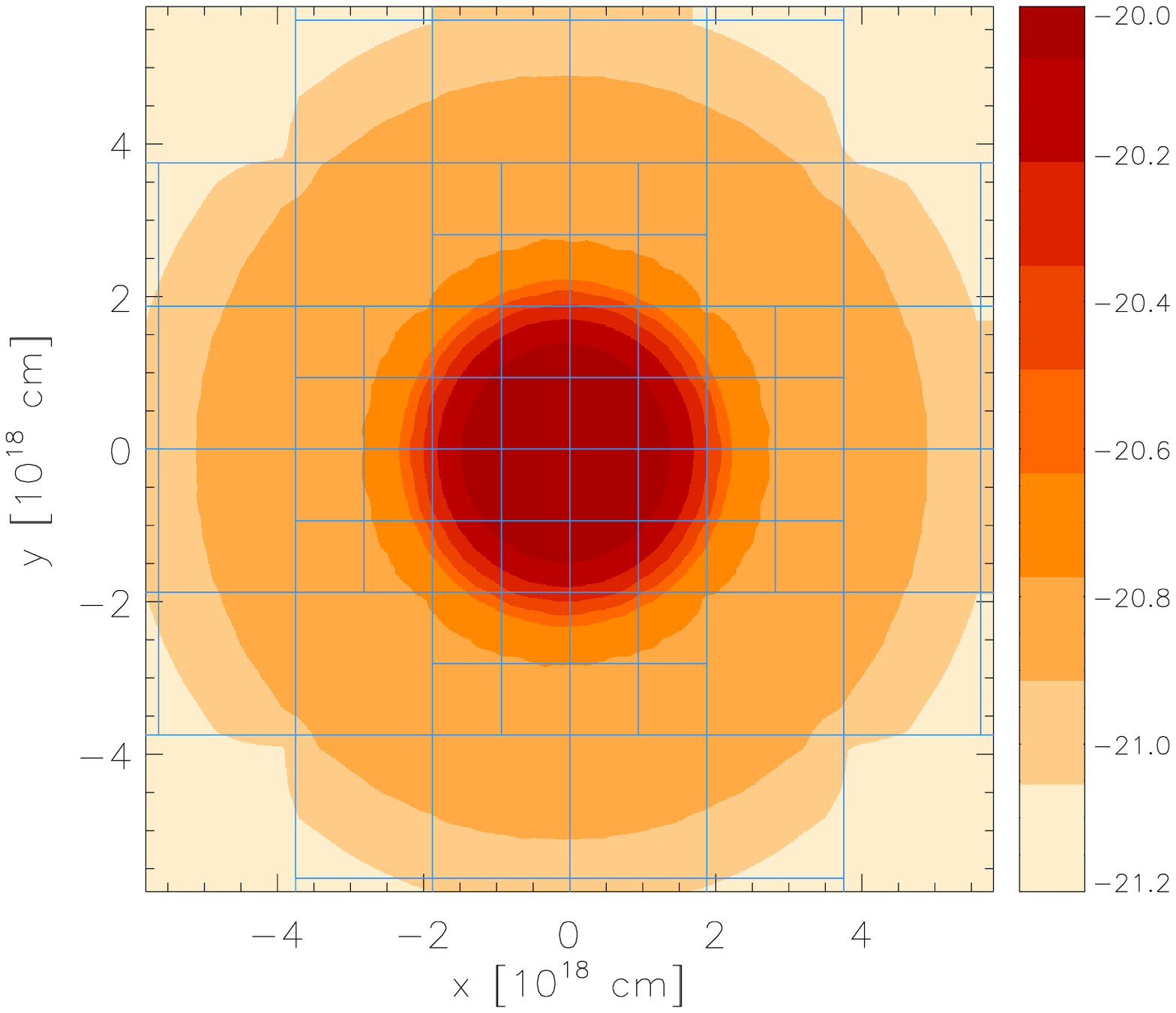} &
\includegraphics[width=84mm]{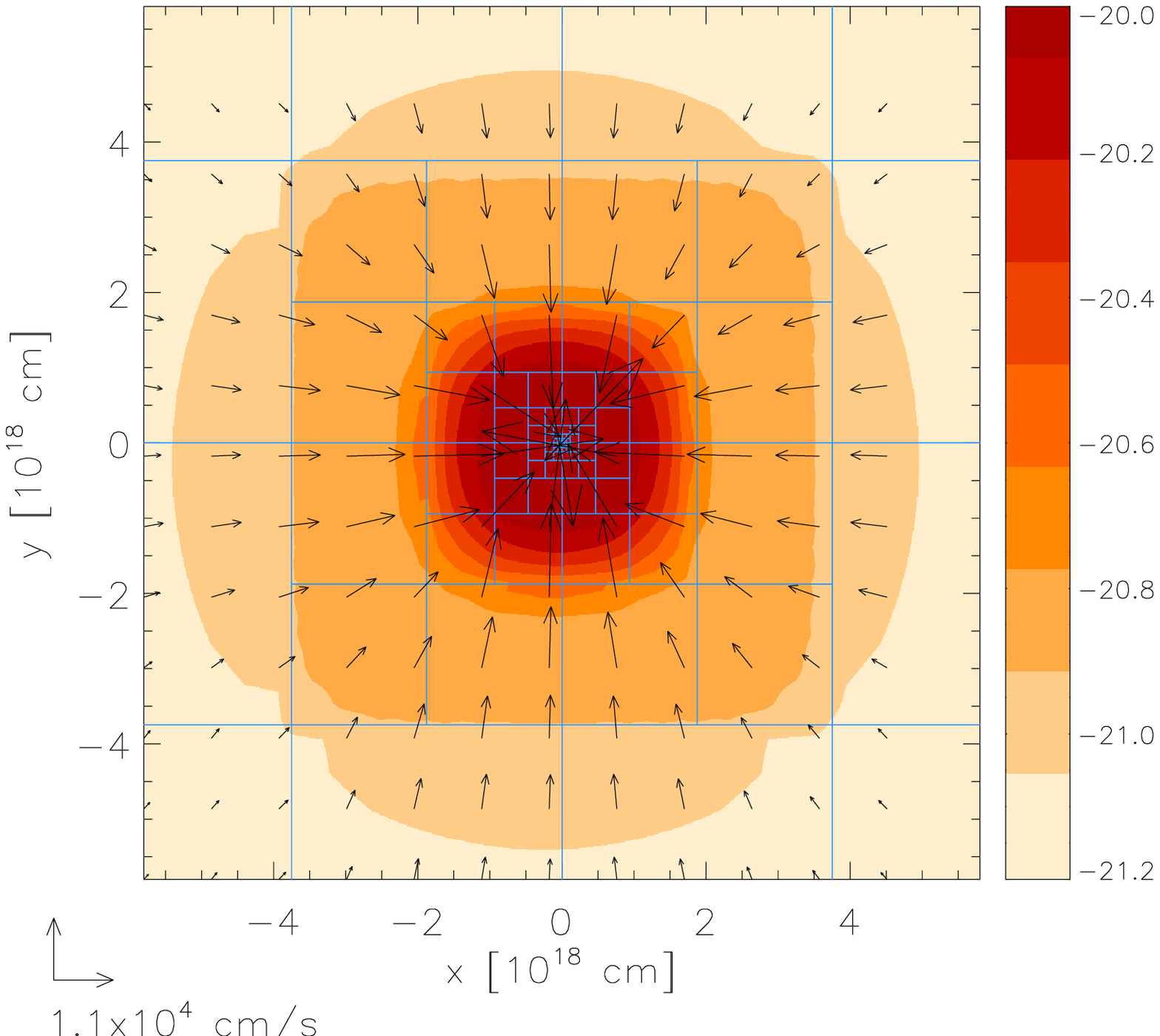} \\
\end{tabular}
\caption{\label{fig:end-state} 
Density pseudo--colours from the end--states of the ideal (left--hand panels) and non-ideal (right--hand panels) simulations of the quasi--static collapse showing the blocks.  Slices are taken to show x--z (top panels) and x--y (bottom panels).  Our ideal end--state is at 11.9 Myr while the non--ideal end--state is at 11.4 Myr.  Densities are restricted to the max/min value in the ideal simulation.  Each block contains 8$^{3}$ computational cells.  Typical velocity vectors, where appropriate, are given in the bottom left corner of each image.    }
\end{figure*}

\begin{figure*}
\includegraphics[width=84mm]{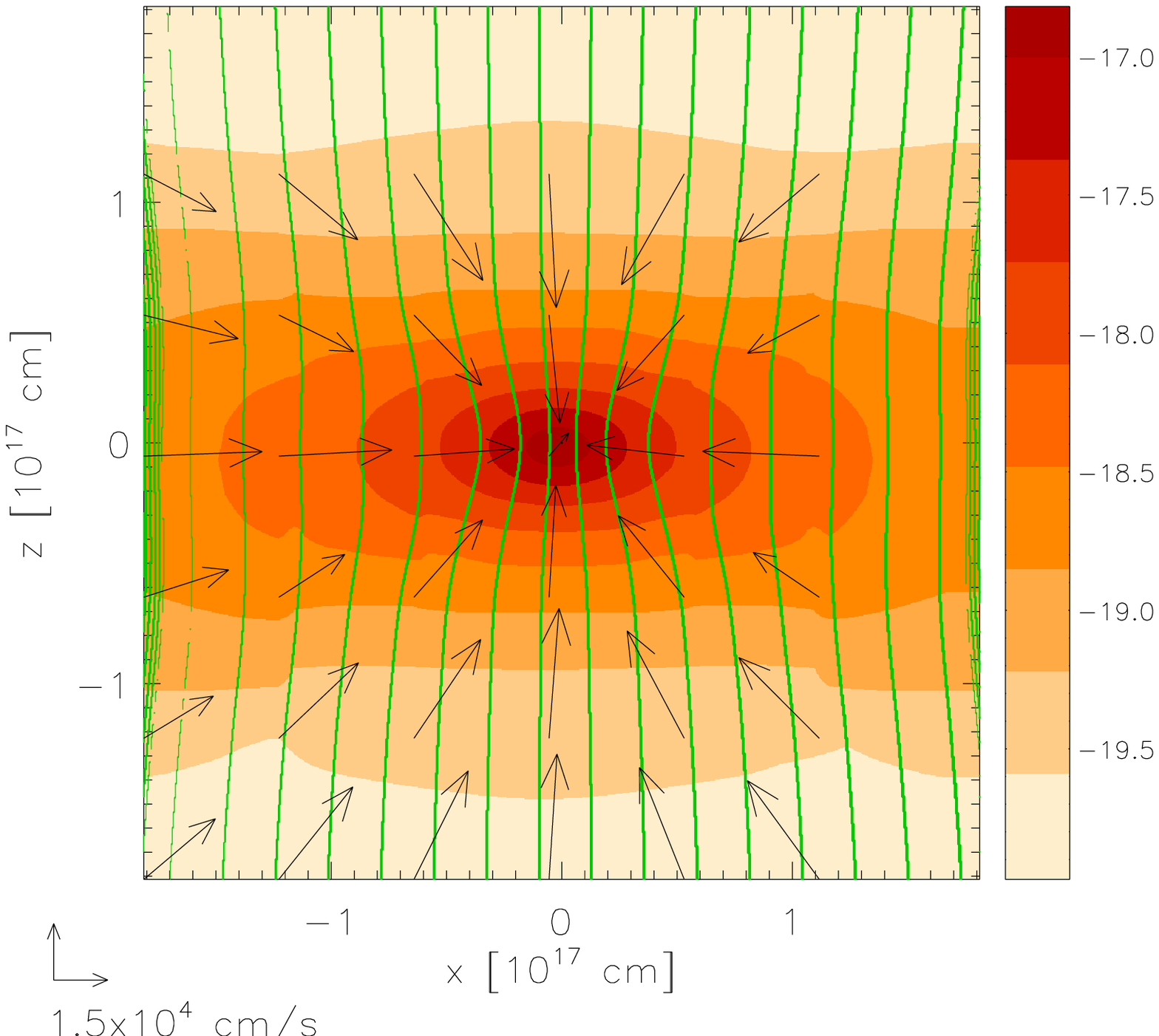}
\caption{\label{fig:field}
Zoomed--in density pseudo--colour from the end--state of the non--ideal quasi--static collapse showing the bending of the magnetic field lines.  Free--fall collapse has begun and accreting gas in the disc pinches the field inwards.  Typical velocity vectors are shown in the bottom left of the image.  
}
\end{figure*}

In Fig.~\ref{fig:end-state} we find that the ideal disc tends to be thicker, in the sense that density distributions are more spread out in the disc.  The non--ideal disc has a significantly higher central density with a corresponding higher degree of refinement (the ideal case has 3 levels of refinement while the non--ideal case has 8 by the time we end the simulation at 11.9 and 11.4 Myr respectively).  This is because the non--ideal disc is more collapsed due to a slight infall motion present of about 0.1 km s$^{-1}$.  Since free--fall collapse has begun in the non--ideal case by our final plot file we can observe that field lines are being pinched in by radially accreting material (shown in Fig.~\ref{fig:field}).  As we only output plot files every 0.1 Myr, we did not follow the collapse further (also cooling restrictions as well as shock and drift heating would become increasingly important in further stages of the clouds evolution).  

Also of note is the preservation of axisymmetry in the simulation, but also the non--linear outline of the thin--disc.  The latter is due in part to the collapsing sphere which falls more quickly in the centre regions than its sides.  This leaves peaks and ripples in the disc's surface as the outer parts of the sphere eventually join the disc.   The disc is also much larger in radius than the initial sphere, aiding our argument that box--size is important in this type of calculation.  

We have correctly found the time--scale of collapse described by \citetalias{FM1993b}, matching all qualitative features of the quasi--static collapse.  We conclude that we have passed this qualitative test.  More importantly, we hope that we have set up a version of the collapse that is easier to reproduce.  However, optimization is required in the parameters in order for this test case to be streamlined for quicker testing.

%%%%%%%%%%%%%%%%%%%%%%%%%%%%%%%%%%%%%%%%%%%%%%%%%%%%%%%%%%%%%%%%%%%%%%%%%%%%%%%%
%%%%%%%%%%%%%%%%%%%%%%%%%%%%%%%%%%%%%%%%%%%%%%%%%%%%%%%%%%%%%%%%%%%%%%%%%%%%%%%%
%%%%%%%%%%%%%%%%%%%%%%%%%%%%%    Conclusions       %%%%%%%%%%%%%%%%%%%%%%%%%%%%%
%%%%%%%%%%%%%%%%%%%%%%%%%%%%%%%%%%%%%%%%%%%%%%%%%%%%%%%%%%%%%%%%%%%%%%%%%%%%%%%%
%%%%%%%%%%%%%%%%%%%%%%%%%%%%%%%%%%%%%%%%%%%%%%%%%%%%%%%%%%%%%%%%%%%%%%%%%%%%%%%%
\section{CONCLUSIONS AND DISCUSSION}
Our results present a properly implemented and highly accurate application of ambipolar diffusion into the {\sc FLASH} AMR MHD code.  Our model is that of a single--fluid (which provides a computational advantage over multi--fluid methods with a necessary chemical model  at the expense of more stringent physical restrictions). We have the benefit of a full non--isothermal energy equation.  

We have presented the first test case of non--isothermal aspects of ambipolar diffusion through the development of the corresponding C--shock analytical solution, with applications to a general chemistry (or no chemistry at all).  We hope this becomes an important test for future codes.

The quasi--static collapse simulation was modified in a more computationally accessible manner.  The added effect of a larger mass reservoir speeds up the creation of the quasi--static \emph{thin disc}, and slightly shortens the magnetically mediated time--scale for collapse (by about 0.5 Myr).  It may also explain the slight differences we found in power law trends (power laws tended to be lower by about 0.1).  
Specifically, we found that the uniform spheres quickly adopt flat--topped, B--E like density profiles with $n(r) \propto r^{-\alpha}$
where $\alpha = (1.51, 1.89, 2.30) $ for our  ideal MHD, AD MHD, and pure  hydro cases respectively.  The ideal MHD sphere remained
supported by the magnetic field (it was initially subcritical, although with a more concentrated density profile).  Both the AD
and hydro cases went into collapse.  We also found the very interesting scaling of the core magnetic field with the maximum
density  $B_c \propto n_{c}^{0.37}$, which is in good agreement with other published results.
We found similar qualitative features not only to the comparison study, but also to studies involving the collapse of Bonner--Ebert Spheres \citep[e.g.][]{BP2007}. Although only meant as a qualitative test, our quasi--static collapse (alongside previously established work) may prove a useful comparison for observations and other types of collapse simulations looking to test persistence of such a process in our current understanding of star formation.  

In our next paper, we use our new code to investigate the collapse of a rotating, magnetized, Bonnor--Ebert sphere in full 3D and with AD included.  

\section*{Acknowledgments}
We are grateful to Robi Banerjee for many useful discussions about the code. 
We wish to thank the anonymous referee for his/her constructive comments.  
One of us (REP) wishes to thank the Kavli Institute for Theoretical Physics (KITP) at 
Santa Barbara as well as the Canadian Institute for Theoretical Astrophysics (CITA) for hospitality and support enjoyed while this
manuscript was being completed.  We also wish to thank Chris McKee, Dick Crutcher, and Telemachos Mouschovias for
interesting conversations. DD was supported by McMaster University and REP by the Natural Sciences and Engineering
Research Council of Canada. We would like to acknowledge the SHARCNET HPC Consortium for the use of its facilities
at McMaster University.
The software used in this work was in part developed by the DOE--supported ASC/Alliances Center for Astrophysical Thermonuclear Flashes at the University of Chicago.
This research was supported in part by the National Science  
Foundation under
Grant No. PHY05-51164

\appendix
%%%%%%%%%%%%%%%%%%%%%%%%%%%%%%%%%%%%%%%%%%%%%%%%%%%%%%%%%%%%%%%%%%%%%%%%%%%%%%%%
%%%%%%%%%%%%%%%%%%%%%%%%%%%%%%%%%%%%%%%%%%%%%%%%%%%%%%%%%%%%%%%%%%%%%%%%%%%%%%%%
%%%%%%%%%%%%%%%%%%%%%    MHD Implementation    %%%%%%%%%%%%%%%%%%%%%%%%%%%%%%%%%
%%%%%%%%%%%%%%%%%%%%%%%%%%%%%%%%%%%%%%%%%%%%%%%%%%%%%%%%%%%%%%%%%%%%%%%%%%%%%%%%
%%%%%%%%%%%%%%%%%%%%%%%%%%%%%%%%%%%%%%%%%%%%%%%%%%%%%%%%%%%%%%%%%%%%%%%%%%%%%%%%
\section{Implementation of new physics into {\sc FLASH}}\label{sec:mhdimp}
Here we describe how we have coded our model of ambipolar diffusion into {\sc FLASH}.

%%%%%%%%%%%%%%%%%%%%%%%%%%%%%%%%%%%%%%%%%%%%%%%%%%%%%%%%%%%%%%%%%%%%%%%%%%%%%%%%
%%%%%%%%%%%%%%%%%%%%%%%%%%%%%%%%%%%%%%%%%%%%%%%%%%%%%%%%%%%%%%%%%%%%%%%%%%%%%%%%
%%%%%%%%%%%%%%%%%%%%%%%%%%%%%%%%    Imp AD    %%%%%%%%%%%%%%%%%%%%%%%%%%%%%%%%%%
%%%%%%%%%%%%%%%%%%%%%%%%%%%%%%%%%%%%%%%%%%%%%%%%%%%%%%%%%%%%%%%%%%%%%%%%%%%%%%%%
%%%%%%%%%%%%%%%%%%%%%%%%%%%%%%%%%%%%%%%%%%%%%%%%%%%%%%%%%%%%%%%%%%%%%%%%%%%%%%%%
\subsection{Implementation of ambipolar diffusion}\label{sec:impad}

To put the equations in code form we will have to first separate sources and fluxes, and then add these to the \textsc{Flux} and \textsc{Source} arrays during the sweeping process of \textsc{mhd\_sweep.f90} in {\sc FLASH2.5}.  This module performs the sweeps, calls the interpolation of the data for ideal terms and calls fluxes and sources for a given sweep direction (including non--ideal data).  After this, it evolves the MHD equations.  In coding the ambipolar diffusion terms we follow closely how the resistive fluxes are applied (see \textsc{mhd\_add\_resistive\_fluxes.f90} in {\sc FLASH2.5}).  A central difference method is used to discretize derivatives ($\bmath{\nabla \times B}$ terms) at cell centres.  This should be an effective method as the ambipolar diffusion terms, like the Ohmic dissipation terms, are parabolic in nature.  They won't require the highly involved interpolation the rest of the MHD equations undergo to be accurate (the ideal MHD equations are hyperbolic).  

To illustrate the central differencing technique, we evaluate an imaginary flux term of the form 
\begin{equation}
F = \left(\frac{A_{i,j,k} B_{i,j,k}}{C_{i,j,k}}\right) + \frac{d D_{i,j,k}}{d x_{i,j,k}} +  \frac{D_{i,j,k}}{d y_{i,j,k}}, 
\end{equation}
where $i$,$j$, and $k$ locate a given cell in the current block and $A$, $B$, $C$, $D$, $x$ and $y$ are evaluated at cell centres.   The x and y coordinates are represented by $x_{i,j,k}$ and $y_{i,j,k}$ respectively (we omit the z coordinate for simplicity).  Let the sweep direction be the x direction, and averaged terms be represented by a line over the character.  Our averaging technique is as follows\footnote{Note that we have also used a similar averaging technique which averages final terms as a whole, rather than basic values (like B$_x$, $\beta_\mathrm{AD}$ or in this imaginary case, $A$).  We find no significant difference between the two.  However, the one presented here more closely resembles that which is already employed in the {\sc FLASH} code.},
\begin{equation}
\overline{A} = 0.5 \left(A_{i,j,k} + A_{i-1,j,k}\right), 
\end{equation}
where $\overline{B}$ and $\overline{C}$ are found in a similar fashion.  Derivatives in line with the sweep follow:
\begin{equation}
\frac{d D}{d x} = \left(\frac{D_{i,j,k} - D_{i-1,j,k}}{x_{i,j,k} - x_{i-1,j,k}}\right).
\end{equation}  
Derivatives that are against the sweep follow this form:
\begin{equation}
\frac{d D}{d y} = 0.5 \left(\frac{\left(D_{i,j+1,k} - D_{i,j-1,k}\right) + \left(D_{i-1,j+1,k} - D_{i-1,j-1,k}\right)}{y_{i,j+1,k} - y_{i,j-1,k}}\right).
\end{equation}
Thus our final flux term will have the form:
\begin{equation}
F = \left(\frac{\overline{A}\ \overline{B}}{\overline{C}}\right)+ \frac{d D}{d x} + \frac{d D}{d y}.
\end{equation}

Now we will evaluate the ambipolar diffusion flux terms and put them in a form that makes sense for computation.  In the code we have applied the above averaging technique.  Please note our matrix notation for flux terms corresponding to $\bmath{B}$.  We have chosen to make each column correspond to $B_x$, $B_y$ and $B_z$ from left to right.  Rows for flux and source vectors or matrices correspond to the direction of the sweep; x direction, y direction and z direction from top to bottom.  For example, during a sweep in the x direction the flux terms only receive additions from the first component in the corresponding vector.  

For simplicity we define the vector,
\begin{equation}\label{Avector}
\begin{split}
\bmath{A} &= -\beta_\mathrm{AD} \bmath{B \times[B \times(\nabla \times B)]}\\
               &= -\beta_\mathrm{AD} \left(
\begin{array}{c}
B_{x}(B_{y} j_{y} + B_{z} j_{z}) - j_{x}(B_{y}^{2} + B_{z}^{2})\\
B_{y}(B_{x} j_{x} + B_{z} j_{z}) - j_{y}(B_{x}^{2} + B_{z}^{2})\\
B_{z}(B_{x} j_{x} + B_{y} j_{y}) - j_{z}(B_{x}^{2} + B_{y}^{2}) 
\end{array}\right), 
\end{split}
\end{equation}    
where $\bmath{j}= \bmath{\nabla \times B}$.  

We find that the flux vectors for the magnetic field components are,
\begin{equation}\label{fluxB}
\left(
\begin{array}{ccc}
\bmath{F}_{B_{x}} & \bmath{F}_{B_{y}} & \bmath{F}_{B_{z}}
\end{array}\right) 
\ +\!=
\left( 
\begin{array}{ccc}
0 & -A_{z} & A_{y}\\
A_{z} & 0 & -A_{x}\\
-A_{y} & A_{x} & 0
\end{array}\right).
\end{equation}

Similarly we define the vector,
\begin{equation}\label{avector}
\begin{split}
\bmath{a} &=  \bmath{(j \times B})\\
               &=  \left(
\begin{array}{c}
j_{y} B_{z} - j_{z} B_{y}\\
j_{z} B_{x} - j_{x} B_{z}\\
j_{x} B_{y} - j_{y} B_{x}
\end{array}\right), 
\end{split}
\end{equation}
for the source.  The sources for \eqref{induction} are coded as follows (only terms with a $\beta_\mathrm{AD}$ need to be added, others are already taken care for):
\begin{equation}\label{sourceB}
S_{B} \ +\!= -\beta_\mathrm{AD} (\bmath{\nabla \cdot B})\left(
\begin{array}{ccc}
a_{x} & 0 & 0\\
0 & a_{y} & 0\\
0 & 0 & a_{z}
\end{array}\right), 
\end{equation}
where we have split up the source into a vector for each direction for simplicity (following our notation outlined above).  

The energy flux works out to,
\begin{equation}\label{fluxE}
\bmath{F}_{E} \ +\!= -\beta_\mathrm{AD} B^{2}\left(
\begin{array}{c}
a_x\\
a_y\\
a_z
\end{array}\right).
\end{equation}

The ambipolar diffusion heating term works out to,
\begin{equation}\label{sourceE}
\bmath{S}_{E} \ +\!= \beta_\mathrm{AD} \left(
\begin{array}{c}
a_x^2\\
a_y^2\\
a_z^2
\end{array}\right), 
\end{equation}
where again we've chosen how to split up the source.

Finally, the $\bmath{\nabla \cdot B}$ source term works out to,
\begin{equation}\label{sourceEdivB}
\bmath{S}_{E} \ +\!= -\beta_\mathrm{AD} \left(
\begin{array}{c}
(B_{x} a_{x})(\bmath{\nabla \cdot B})\\
(B_{y} a_{y})(\bmath{\nabla \cdot B})\\
(B_{z} a_{z})(\bmath{\nabla \cdot B})
\end{array}\right),
\end{equation}
where we've chosen to split the source term up partly to save coding and match the usage of $\bmath{a}$'s components in Equation \eqref{sourceB}.  The energy equations in this form are quite efficient, relying on the calculation of only one component of $\bmath{a}$ for each sweep.

%%%%%%%%%%%%%%%%%%%%%%%%%%%%%%%%%%%%%%%%%%%%%%%%%%%%%%%%%%%%%%%%%%%%%%%%%%%%%%%%
%%%%%%%%%%%%%%%%%%%%%%%%%%%%%%%%%%%%%%%%%%%%%%%%%%%%%%%%%%%%%%%%%%%%%%%%%%%%%%%%
%%%%%%%%%%%%%%%%%%%%%%%%%%%%%    UNITS    %%%%%%%%%%%%%%%%%%%%%%%%%%%%%%%%%%%%%%
%%%%%%%%%%%%%%%%%%%%%%%%%%%%%%%%%%%%%%%%%%%%%%%%%%%%%%%%%%%%%%%%%%%%%%%%%%%%%%%%
%%%%%%%%%%%%%%%%%%%%%%%%%%%%%%%%%%%%%%%%%%%%%%%%%%%%%%%%%%%%%%%%%%%%%%%%%%%%%%%%
\subsection{Units, constants and dimensionality}\label{sec:units}
One advantage of {\sc FLASH} is that the dimensionless constants are all $1.0$, save the constant for the magnetic field.  The four principle scaling constants are \citep{1999JCP.154.284P}:
\begin{equation}\label{dimless}
\begin{split}
a_{\infty} = &\ 1.0\ \left[\frac{\mathrm{cm}}{\mathrm{s}}\right]\\
\rho_{\infty} = &\ 1.0\ \left[\frac{\mathrm{g}}{\mathrm{cm}^{3}}\right]\\
L = &\ 1.0\ [\mathrm{cm}]\\
\mu_{0} = &\ 4 \pi \ \left[\frac{\mathrm{cm}\ \mathrm{s}^{2}\ \mathrm{G}^{2}}{\mathrm{g}}\right], 
\end{split}
\end{equation}
from which most variables can be made dimensionless through combinations of these constants.

Important examples are the magnetic field, $B' = (a_{\infty} \sqrt{\rho_{\infty} \mu_{0}})^{-1} B$ and $\beta_\mathrm{AD}' = a_{\infty} \rho_{\infty} L^{-1} \mu_{0} \beta_\mathrm{AD}$.  Otherwise, all other variables are quite straightforward.  One can imagine equations (\ref{continuity}--\ref{energy}) the same, but drop all $\mu_{0}$ terms to $1$.  

%%%%%%%%%%%%%%%%%%%%%%%%%%%%%%%%%%%%%%%%%%%%%%%%%%%%%%%%%%%%%%%%%%%%%%%%%%%%%%%%
%%%%%%%%%%%%%%%%%%%%%%%%%%%%%%%%%%%%%%%%%%%%%%%%%%%%%%%%%%%%%%%%%%%%%%%%%%%%%%%%
%%%%%%%%%%%%%%%%%%%%%%%%%%%%%    Timesteps    %%%%%%%%%%%%%%%%%%%%%%%%%%%%%%%%%%
%%%%%%%%%%%%%%%%%%%%%%%%%%%%%%%%%%%%%%%%%%%%%%%%%%%%%%%%%%%%%%%%%%%%%%%%%%%%%%%%
%%%%%%%%%%%%%%%%%%%%%%%%%%%%%%%%%%%%%%%%%%%%%%%%%%%%%%%%%%%%%%%%%%%%%%%%%%%%%%%%
\subsection{Timesteps}\label{sec:tad}

The timestep is taken from equation \eqref{adtime--scale} and written as:
\begin{equation}
\begin{split}
\tau_{\mathrm{AD}} = & \ T_0 \frac{(\Delta x)^{2}}{\eta_{\mathrm{AD}}}\\
 = &\ T_0\left[\frac{(x_{i,j,k}-x_{i-1,j,k})^{2}}{\overline{\beta_\mathrm{AD}}\ \overline{B}^2 }\right], 
\end{split}
\end{equation}
where $T_0$ is chosen as  $\frac{1}{120}$ in our runs (as used in \citet{1995ApJ...442..726M}).

The $\Delta x$ represents the coordinate of the sweep, so it would be $\Delta z$ if the sweep was in the z direction.  The smallest such timestep in a sweep is given to the timestep chooser which multiplies the smallest timestep from all processes by a CFL factor of typically $0.8$.  The ambipolar diffusion timestep is small and usually dominates over other timesteps (such as from the cooling or ideal MHD processes).

%%%%%%%%%%%%%%%%%%%%%%%%%%%%%%%%%%%%%%%%%%%%%%%%%%%%%%%%%%%%%%%%%%%%%%%%%%%%%%%%
%%%%%%%%%%%%%%%%%%%%%%%%%%%%%%%%%%%%%%%%%%%%%%%%%%%%%%%%%%%%%%%%%%%%%%%%%%%%%%%%
%%%%%%%%%%%%%%%%%%%%%%%%%%%%%    Bibliography %%%%%%%%%%%%%%%%%%%%%%%%%%%%%%%%%%
%%%%%%%%%%%%%%%%%%%%%%%%%%%%%%%%%%%%%%%%%%%%%%%%%%%%%%%%%%%%%%%%%%%%%%%%%%%%%%%%
%%%%%%%%%%%%%%%%%%%%%%%%%%%%%%%%%%%%%%%%%%%%%%%%%%%%%%%%%%%%%%%%%%%%%%%%%%%%%%%%

\bibliographystyle{mn}
\bibliography{paper}

\label{lastpage}

\end{document}